\documentclass{aa} 

\usepackage[nolist]{acronym}

\usepackage{graphicx}
\usepackage{threeparttablex}
\usepackage{txfonts}

\usepackage{amsmath,amsfonts,amssymb}
\usepackage{bm}   
\usepackage{graphicx}
\usepackage[colorlinks=true, allcolors=blue]{hyperref}
\usepackage{astro_bib_macro}
\usepackage{ulem}

\DeclareMathOperator{\sinc}{sinc}

\usepackage{booktabs}

\usepackage{listings}

\graphicspath{{./}{./figures/}}   

\makeatletter
\renewcommand*\aa@pageof{, page \thepage{} of \pageref*{LastPage}}
\makeatother

\begin{document}

\title{Extended Linearity in the High-Order Wavefront Sensor for the Roman Coronagraph}

\titlerunning{THD2 Roman PW Probes}
\authorrunning{Laginja {et al.}}

\author{Iva Laginja\inst{\ref{inst-a},}\inst{\ref{inst-b}} \and Pierre Baudoz\inst{\ref{inst-b}} \and Johan Mazoyer\inst{\ref{inst-b}} \and Axel Potier\inst{\ref{inst-b}} \and Raphaël Galicher\inst{\ref{inst-b}}\and Faouzi Boussaha\inst{\ref{inst-c}} }

\institute{NOVA/Leiden University, Einsteinweg 55, 2333 CC Leiden, The Netherlands\label{inst-a}
\and
LIRA, Observatoire de Paris, Université PSL, Sorbonne Université, Université Paris Cité, CY Cergy Paris Université, 92190 Meudon, France\label{inst-b}
\and
LUX, Observatoire de Paris, Université PSL, Université PSL, Sorbonne Université, CNRS, 
61 avenue de l’Observatoire, 75014 Paris, France\label{inst-c}\\
\email{iva.laginja@obspm.fr}
}

\date{Received 15 January 2025; accepted 30 March 2025}

  \abstract
  {The Coronagraphic Instrument (CGI) aboard the Roman Space Telescope aims to achieve unprecedented levels of contrast for direct imaging of exoplanets, serving as a critical technology demonstrator for future missions like the Habitable Worlds Observatory (HWO). Achieving these goals requires advanced wavefront sensing and control (WFS\&C) strategies, including the use of pair-wise (PW) probing to estimate the electric field in the focal plane. The optimization of PW probe designs is vital to enhance performance and reduce operational overhead.}%
  {
    This study investigates the performance of different probe designs for PW probing in the context of Roman CGI. Specifically, we aim to compare the classic sinc-sinc-sine probes, previously introduced single-actuator probes, and newly proposed sharp sinc probes in terms of their effectiveness in focal-plane modulation, resilience to non-linearities at high probe amplitudes, and overall impact on the convergence and contrast levels achieved in laboratory demonstrations.
  }
   {
   We conducted experiments on the THD2 testbed, configured to simulate Roman CGI with a custom-made Hybrid Lyot Coronagraph (HLC). We evaluated the three probe designs through closed-loop WFS\&C experiments using PW probing for electric field estimation and electric field conjugation (EFC) for wavefront correction. Simulations and hardware tests assessed contrast convergence and the impact of non-linear terms at varying probe amplitudes. We also explored low-flux scenarios to demonstrate the effectiveness of high-amplitude probes in reducing exposure times or closing the loop on faint targets.
   }
   {
   Single-actuator probes emerged as the most effective design, offering faster convergence and reduced susceptibility to non-linear effects at high amplitudes compared to sinc-sinc-sine probes. Sharp sinc probes performed moderately well but were less robust than single-actuator probes. High-amplitude single-actuator probes demonstrated advantages in dark-hole digging under low-flux conditions, achieving faster iterations without significant degradation in contrast performance. The THD2 testbed, operating in a contrast regime analogous to Roman CGI, validated these results and underscored its role as a critical platform for advancing WFS\&C techniques.
   }
   {}
   

   \maketitle
%

\section{Introduction}
\label{sec:introduction}

Aboard the Roman Space Telescope \citep{Perkins2024RomanSpaceTelescope}, the Coronagraphic Instrument (CGI) will be a major milestone towards the technology required for the direct imaging of Earth-like exoplanets around Sun-like stars. The mission goal is to demonstrate key technologies for high-contrast imaging (HCI), including deformable mirrors (DMs). It is predicted to observe targets at a raw contrast of $10^{-8}$ or better in a high-contrast detector area (``dark hole'', DH) from 3--9~$\lambda/D$ \citep{Bailey2023NancyGraceRomanSpace, Kasdin2020NancyGraceRoman, Noecker2016CoronagraphInstrument}, with $\lambda$ the observing wavelength and $D=2.4$~m the diameter of the telescope aperture. While this is magnitudes better than the currently achievable coronagraphic performance from the ground and from space alike, it is still more than an order of magnitude less capable than what is required for the next NASA flagship mission, the Habitable Worlds Observatory \citep[HWO,][]{Feinberg2024TheHabitableWorlds, OMeara2024TheHabitableWorlds}, which aims to achieve a raw contrast of $\sim10^{-10}$. As a pathway towards this goal, CGI will be the first space-based instrument to deploy a pair of DMs for active sensing and control of the coronagraphic electric field. A coronagraph suppresses the light of the observed on-axis star while altering the faint light of a faint companion as little as possible. While there exists today a long list of very performant coronagraphs in theory, they are very sensitive to aberrations, which necessitates the usage of wavefront sensing \& control (WFS\&C) strategies.

The algorithms that will be used in these WFS\&C loops have been extensively tested in the laboratory facilities of NASA's Jet Propulsion Laboratory (JPL). In 2024, CGI underwent testing in a thermal vacuum chamber (TVAC) tests and successfully ran the control loops in flight-like conditions \citep{Cady2025HighOrderWavefront}. Nevertheless, the WFS\&C implementation for the mission has been set prior to significant developments in the field. This means that some algorithmic choices, while sufficient, might not be optimal if the ultimate goal is to peruse CGI results for the developments towards HWO.

With CGI being a technology demonstration mission, it has a minimal set of goals to meet with a strict baseline of instrumental modes, and very limited observing time. This provides an opportunity to explore technological improvements that are directly compatible with the current HCI implementation of CGI. In this paper, we explore alternatives in the DM commands to be used for the probing of the electric field. The resulting estimates are used to determine the DM commands for wavefront control that iteratively improve the detected raw contrast. Updating the baseline probe files represents a very non-invasive change to the instrument operations that would allow us to explore new algorithm variants while staying easily within Roman mission constraints.

In this paper, we are not aiming to achieve the best possible contrast, nor are we focusing on refining control strategies. Instead, our focus is specifically on the estimation process within the WFS\&C framework, and in monochromatic light. By isolating the effect of DM probes, we aim to evaluate and improve the accuracy and efficiency of electric field estimation in the Roman CGI scheme.

In Sec.~\ref{sec:hlc_on_roman}, we briefly describe the Hybrid Lyot Coronagraph (HLC) of Roman that serves as the CGI coronagraph we refer our study to, we introduce its planned WFS\&C strategy, and results of ground tests. We provide a brief summary of the pair-wise (PW) estimation algorithm as prepared for CGI and explain how it compares to recent developments and the current state of the field. In Sec.~\ref{sec:roman_hlc_thd2}, we give a description of the THD2 testbed at LIRA (Observatoire de Paris - PSL) in Meudon, France. We present the optical masks used to emulate Roman CGI, and the design and manufacturing process for the testbed's HLC masks. In Sec.~\ref{sec:probe_choice}, we repeat the considerations for the original probes that were developed for PW probing, recall the interest of single-actuator probes and introduce the new sharp sinc probes. We also compare the closed-loop performance of these three types of probes in a standard WFS\&C setup in numerical simulations and on hardware. In Sec.~\ref{sec:probe_linearity}, we explore the closed-loop performance of the same probes at high probe amplitudes on hardware, and numerically investigate the behavior of non-linear terms of the electric field expansion, comparing between the classic sinc and single-actuator probes. We then introduce an application case for high-amplitude single-actuator probes, which allow to significantly speed up the WFS\&C loop without violating the linear field expansion. Finally, in Sec.~\ref{sec:discussion} we discuss the results of this paper and find a conclusion in Sec.~\ref{sec:conclusions}.

\section{Roman Coronagraph and WFS\&C strategy}
\label{sec:hlc_on_roman}

\subsection{Hybrid Lyot Coronagraph on Roman}
\label{subsec:hybrid_lyot_coronagraph}

Roman CGI includes several required and optional coronagraphs designed to be tested during the mission lifetime \citep{Riggs2021FlightMaskDesigns} in conjunction with four distinct, large color bandwidths 1--4 (central wavelengths 575~nm, 660~nm, 760~nm and 825~nm), each with a list of subbands. The primary required coronagraph modes are an HLC, which operates within a narrow field of view (FOV) of 3--9~$\lambda/D$, and a wide-FOV Shaped Pupil Coronagraph (SPC), both operating in band one. They have been extensively simulated and modeled for the mission \citep{Krist2023EndToEndNumericalModeling, Krist2015OverviewWFIRSTAFTA}. Beyond these, three other coronagraph configurations, all of which are SPCs, have been designated as ``best effort'' modes. These coronagraphs are intended to enable both spectroscopy and polarimetry.

The main coronagraph considered in this paper is the HLC \citep{Trauger2012ComplexApodizationLyot, Trauger2011AHybridLyotCoronagraph, Moody2008DesignANdDemonstrationOfHybrid, Sidick2007BehaviorOfImperfectBand, Kuchner2002ACoronagraphWithABandLimited}. It is based on the classical Lyot coronagraph \citep[CLC,][]{Lyot1932EtudeCouronneSolaire,Vilas1987CoronagraphAstronomicalImaging}. While the CLC uses an entirely opaque focal-plane mask (FPM) with a diameter of several resolution elements to block the star light, the HLC instead uses a similarly sized spot that alters both phase and amplitude of the stellar electric field. The concrete design of the FPM for the HLC on Roman consists of two key layers: a metallic layer (nickel) and a dielectric layer (MgF$_2$ or cryolite), both deposited onto a fused silica substrate \citep{Trauger2016HybridLyotCoronagraphFor, Trauger2013ComplexApodizedLyot}. The metallic layer primarily controls the attenuation of incoming starlight, blocking most of it, while the dielectric layer modulates the phase of the remaining light that passes through the mask. This allows the HLC to suppress diffracted light more effectively, achieving the DH for HCI. The dielectric layer's thickness is carefully profiled to adjust the phase of the light, while the FPM overall has an intensity transmittance of approximately $10^{-4}$. Additionally, a dimple with a diameter of 1.22~$\lambda/D$ is incorporated into the dielectric layer to introduce a $\pi/2$ phase shift, which is used to improve low-order wavefront sensing by reflecting a portion of the starlight to a wavefront sensor, crucial for stabilizing pointing jitter \citep{Shi2016LowOrderWavefrontSensing}.

\subsection{WFS\&C strategy for Roman CGI, and demonstrations}

Compensating wavefront errors (WFEs) is an iterative, two-step process. First, a wavefront sensor captures the information about the current state of the electric field and aberrations contained therein. This is used as the input to the second step, in which a control algorithm computes the correction to be applied to the system and, in the case of HCI, improves the contrast in the DH.

For active WFS\&C, CGI will run several concurrent control loops that allow to stabilize the DH on-sky. The Line of Sight (LoS) Control Loop will perform fast control of the tip and tilt modes, while a dedicated Focus Control Loop and Zernike Control Loop will perform slow control of Zernike modes Z4 (focus) through Z11 (spherical). All three of these rely on wavefront measurements by a dedicated low-order wavefront sensor (LOWFS) that obtains light from a reflection off of the back of the FPM of the HLC. Meanwhile, the DH will be created with the HOWFSC which senses the electric field directly in the focal plane of the detector, to minimize non-common path errors. While LoS corrections and LOWFS results will be calculated on board, the HOWFSC loop will be run with ``ground-in-the-loop''. This means that the recorded focal-plane images for the wavefront sensing will be sent to a ground station where the estimate and the control step will be computed before sending it back to the telescope.

The baseline strategy for the HOWFSC loop is electric field estimation with pair-wise (PW) probing \citep{Give'on2011PairwiseDeformableMirror}, using three pairs of probes of opposite signs, and an unmodulated image to monitor progress. Correction is done with the electric field conjugation (EFC) controller \citep{Give'on2007ClosedLoopDM}. EFC models the field near the current position and optimizes the DM actuator settings to minimize the electric field over the discrete DH pixels using a least-squares solver. The CGI extensions to the basic EFC approach include several enhancements, like pixel and actuator weighting to remove dead pixels and dead or tied actuators, and to emphasize specific areas in the focal plane \citep{Thomas2010LaboratoryTestOfApplication}, for example to get a better estimation and control in pixels closer to the optical axis. It also uses electric field estimates from different wavelengths to capture chromatic variations, and changes the regularization term as a function of iteration, which has proven crucial to achieve best contrast \citep[``$\beta$ bumping'',][]{Cady2017ShapedPupilCoronagraphy, Seo2017HybridLyotCoronagraph, Marx2017ElectricFieldConjugation}.

High contrast around $10^{-9}$ inside the DH was demonstrated with the HLC in a controlled testbed environment in both narrow-band and broadband light \citep{Seo2017HybridLyotCoronagraph}, while also addressing key technological challenges such as low-order wavefront sensing and control, and line-of-sight disturbances. The coronagraph was further tested in more realistic, flight-like conditions, where WFC was performed under time constraints and simulations of ground-to-orbit changes. These laboratory validations showed that the HLC is in principle able to reach the $10^{-7}$ contrast requirement by a comfortable margin of almost two orders of magnitude \citep{Zhou2020RomanCGItedtbedHOWFSC, Zhou2019WFIRSTPhaseHLC, Seo2019TestbedDemonstrationHighcontrast, Seo2018HybridLyotCoronagraph, Seo2017HybridLyotCoronagraph, Seo2016HybridLyotCoronagraph}.

During TVAC tests in March and April of 2024, the HLC was used in closed-loop HOWFSC operating in imaging band 1. As per mission requirements, it needed to achieve a mean contrast better than $5 \times 10^{-8}$ for coherent light and $10^{-7}$ for incoherent light in a 360$^{\circ}$ DH region spanning 6--9 $\lambda/D$, even if the HLC typically operates within 3--9 $\lambda/D$. In this context, coherent light is any light that is well-modulated by the PW probes and thus ``seen'' by the estimator, while incoherent light is the difference of an unprobed image with the estimated coherent intensity.
During these near-FOV tests, the HLC reached a contrast of $0.98 \times 10^{-8}$ in coherent and $2.35\times 10^{-8}$ in incoherent light measured from 6--9 $\lambda/D$, thus meeting requirements \citep{Cady2025HighOrderWavefront}. Moreover, the achieved contrast was constrained by the available testing time rather than any known systematic effect. This means that CGI could well reach the predicted contrast level of several $10^{-9}$ once in orbit \citep{Zhou2023RomanCoronagraphHOWFSC}.

\subsection{Summary of pairwise probing}

In the following, we briefly recall the formalism of PW probing. We introduce a simple formalism and we use some of the equations presented in \citep{Groff2016MethodsLimitationsFocal}.

We call $E_{\mathrm{pup}}(u,v)$ the electric field in the pupil plane that propagates through the coronagraph to give the image-plane electric field~$E_{\mathrm{im}}(x,y)$. It can be expressed using the linear coronagraph operator~$\mathcal{C}$:
\begin{equation}
E_{\mathrm{im}}(x,y) = \mathcal{C} \{ E_{\mathrm{pup}}(u,v) \}.
\label{eq:Eim_C_Epup}
\end{equation}
For PW probing, we add a small known phase, $\Psi_j(u,v)$, the probe, in the pupil plane using the DM. To first order, the electric field in the pupil plane that includes the probe $j$, $E_{\mathrm{pup},\,j}(u,v)$, writes as:
\begin{equation}
    E_{\mathrm{pup},\,j}(u,v) = E_{\mathrm{pup}}(u,v) \,\exp{\left[i\,\Psi_j(u,v)\right]}\simeq E_{\mathrm{pup}}(u,v)\,\left[1+i\,\Psi_j(u,v)\right],
    \label{eq:E_pup_expansion}
\end{equation}
and the image-plane electric field, $E_{\mathrm{im},\,j}$, associated to the probe~$\Psi_j$ is then:
\begin{equation}
     E_{\mathrm{im},\,j}(x,y) \simeq E_{\mathrm{im}}(x,y) + i\,\mathcal{C}\left[E_{\mathrm{pup}}(u,v)\,\Psi_j(u,v)\right].
    \label{eq:E_im}
\end{equation}
The only measurable quantity in Eq.~\ref{eq:E_im} is the image-field intensity~$I_{\mathrm{im},\,j}(x,y) = \left|E_{\mathrm{im},\,j}(x,y)\right|^2$. An unambiguous estimation of the field, $E_{\mathrm{im}}(x,y)$, requires at least two probe pairs per sub-band. In general, this is achieved by applying the probe command $\pm\Psi_j(u,v)$ with opposite signs. The total number of such probe \textit{pairs}, $N_{probe}$, builds a probe \textit{set}, and in this paper we always work with $N_{probe}=3$ pairs of probes in one set.

For the PW estimation, the intensity difference between a positively and negatively probed image, $\Delta I_{\mathrm{im},\,j}(x,y)$, is recorded for all probe pairs in a set. This intensity difference is thus the measured quantity in PW probing. Calculating the probed intensity difference images with $\Delta I_{\mathrm{im},\,j} = |E_{\mathrm{im},\,\psi_j}|^2 - |E_{\mathrm{im},\,-\psi_j}|^2$ from the linearly approximated Eq.~\ref{eq:E_im}, and after dropping the explicit writing of coordinates, we obtain for the PW intensity difference:
    \begin{equation}
    \Delta I_{\mathrm{im},\,j} \simeq 4\ \Re{\left\{i\,E_{\mathrm{im}}\, \mathcal{C}\left[E_{\mathrm{pup}}\,\psi_j\right]^*\right\}},
    \label{eq:PW-delta_i_b}
    \end{equation}
where $\Re$ is the real part and~$^*$ the complex conjugate.

\citet{Groff2016MethodsLimitationsFocal} reshape the 2D electric and intensity fields as vectors, and introduce an interaction matrix, $G$. We call $\mathbf{e}_{\mathrm{im}}$ and $\Delta \mathbf{i}_{\mathrm{j}}$ the vectors of the electric fields~$E_{\mathrm{im}}$ that we are looking for and~$\Delta I_{\mathrm{im,\,j}}$ that we are measuring, respectively. \citet{Groff2016MethodsLimitationsFocal} write:
    \begin{equation}
    \Delta \mathbf{i}_{j} \simeq 4\ \Re{\{\mathbf{e}_{\mathrm{im}} \circ G \bm{\psi}_j\}},
    \label{eq:PW-delta_i_b_Groff}
    \end{equation}
where $\circ$ denotes the complex inner product and $\bm{\psi}_j$ the vector that contains all the actuator commands to create the phase~$\Psi_j$. In this formalism, each row of~$G$ represents the image-plane electric field when pushing one of the DM~actuators.

As shown in \cite{Groff2016MethodsLimitationsFocal}, Eq.~\ref{eq:PW-delta_i_b_Groff} can further be rewritten to:
    \begin{equation}
    \begin{pmatrix} \Delta \mathbf{i}_1 \\ \Delta \mathbf{i}_2 \\  \Delta \mathbf{i}_3 \end{pmatrix} = 4 \begin{pmatrix} \Re{\{G \bm{\psi}_1\}} & \Im{\{G \bm{\psi}_1\}} \\ \Re{\{G \bm{\psi}_2\}} & \Im{\{G \bm{\psi}_2\}} \\ \Re{\{G \bm{\psi}_3\}} & \Im{\{G \bm{\psi}_3\}} \end{pmatrix}   \begin{pmatrix} \Re{\{\mathbf{e}_{\mathrm{im}}\}} \\ \Im{\{\mathbf{e}_{\mathrm{im}}\}} \end{pmatrix},
    \label{eq:matrix-j2}
    \end{equation}
assuming a set of three probe pairs, and $\Im$ taking the imaginary part. The observed delta intensities, $\Delta \mathbf{i}_j$, form the observation vector $\mathbf{z}$, and the state of the electric field forms the state vector $\mathbf{x}$. The observation and state are then related by the observation matrix, or PW matrix, $H$:
    \begin{equation}
    \mathbf{z} = H \mathbf{x}.
    \label{eq:PW-vector-equation_b}
    \end{equation}
To obtain an electric field estimate, $\mathbf{\Hat{x}}$, it is thus necessary to invert the observation matrix, $H$. This is done by taking the pseudo inverse of $H$:
    \begin{equation}
    \mathbf{\Hat{x}} = (H^T\,H)^{-1} H^T\,\mathbf{z} = H^{\dagger}\,\mathbf{z}.
    \label{eq:PW-x-hat-estimate_b}
    \end{equation}
This means that the electric field can be retrieved using:
    \begin{equation}
    \begin{pmatrix} \Re{\{\mathbf{e}_{\mathrm{im}}\}} \\ \Im{\{\mathbf{e}_{\mathrm{im}}\}} \end{pmatrix}= H^{\dagger}
    \begin{pmatrix} \Delta \mathbf{i}_1 \\ \Delta \mathbf{i}_2 \\  \Delta \mathbf{i}_3 \end{pmatrix}.
    \label{eq:matrix-j2_inv_b}
    \end{equation}
    
Equation \ref{eq:PW-x-hat-estimate_b} requires the observation matrix, $H$, to be well-conditioned, which is expressed by the necessary condition for at least two probes $m$ and $n$, in each pixel of the image:
    \begin{equation}
    \Re(G\,\bm{\psi}_m) \Im(G\,\bm{\psi}_n) - \Re(G\,\bm{\psi}_n) \Im(G\,\bm{\psi}_m) \neq 0.
    \label{eq:necessary-condition_b}
    \end{equation}
A very simple way to ensure the condition in Eq.~\ref{eq:necessary-condition_b} is met is to add more and more probes to a set until all pixels in the focal plane are modulated sufficiently so that the condition number of $H$ is low enough. Otherwise, the matrix inversion will be very sensitive to the noise from the given probes. However, any additional probe pair added to a probing set requires two more intensity images to be taken, which increases the time spent on overheads such as creating a DH. It is thus immediately obvious that the much better way is to choose probes that (1) provide a focal-plane modulation as uniform as possible, in as few pairs as possible, and that these probes should (2) introduce as little noise as possible.

A major bias term from probes in a linearly approximated electric field model like in Eq.~\ref{eq:E_im} are their non-linear contributions to the intensity difference, $\Delta \mathbf{i}_{j}$. This is particularly relevant when working with high-amplitude probes, as discussed in \cite{Groff2016MethodsLimitationsFocal}, because it increases the non-linear term levels with respect to the linear approximation. A more detailed discussion of this limitation is given in Sec.~\ref{subsec:nonlinear_terms_PW_observable}.

\subsection{State of the field and recent developments}
\label{subsec:state_of_the_field}

PW estimation has become a cornerstone technique for focal-plane wavefront sensing for small WFE, offering a method to reconstruct the electric field without requiring dedicated hardware such as a Zernike wavefront sensor \citep{NDiaye2013CalibrationOfQuasi,Wallace2011PhaseShiftingZernike}, or a self-coherent camera \citep[SCC,][]{Baudoz2005TheSelfCoherentCamera}.

Since the parallel introduction of EFC and PW probing by \cite{Give'on2007ClosedLoopDM}, progress has been made in particular in refining overall closed-loop performance. Key areas of development include better characterization of DMs, as uncertainties in DM response remain a dominant error source, and the exploration of alternative control strategies such as stroke minimization \citep{Pueyo2009OptimalDarkHole}, Kalman filtering \citep{Groff2013KalmanFilteringTechniques} and Jacobian-free optimization \citep{Will2021JacobianFreeCoronagraphic}.

In terms of probe design for PW probing, a notable shift occurred with \cite{Potier2020ComparingFocalPlane}, who introduced single-actuator probes as an alternative to the sinc-sinc-sine probes first proposed by \cite{Give'on2011PairwiseDeformableMirror}. These single-actuator probes have since been adopted by a number of projects, including HiCAT \citep{Soummer2024HiCAT} and SCExAO \citep{Ahn2021SCExAOTestbedDeveloping}, as well as SPHERE \citep{Potier2022IncreasingTheRawContrast} and MagAO-X \citep{Kueny2024MagaoxPhaseIIUpgrades} on-sky operations. Other probe shape versions include power-law probes \citep{vanGorkom2024TheSpaceCoronagraphOptical}, probes optimized with the control matrix \citep{Will2021WavefrontControlAlgorithmic}, and probes based on pure sine and cosine combinations \citep{Milani2024ModelingAndPerformanceAnalysis,Nishikawa2022TheCoherentDifferential}. However, despite their growing popularity, no systematic comparison of the performance of different probe shapes has been conducted to date.

With future missions like the Roman and HWO relying on focal-plane wavefront sensing -- and PW probing as the most mature technique for this purpose -- it is critical to assess how changes in probe strategies might impact closed-loop performance. Understanding these implications will help optimize operational strategies for Roman and inform the design of next-generation instruments.


\section{THD2 testbed with Roman CGI configuration}
\label{sec:roman_hlc_thd2}

The THD2 (``très haute dynamique'', French for ``very high contrast'') testbed is a high-contrast imaging testbed developed at LIRA, Observatoire de Paris for testing imaging solutions in the visible and near-infrared with achievable contrast levels of $10^{-8}$ to $10^{-9}$  \citep{Baudoz2024PolarizationEffects}.
The testbed uses a multi-DM configuration to correct both phase and amplitude aberrations \citep{Baudoz2018OptimizationPerformanceMultideformable, Baudoz2018StatusPerformanceTHD2}.
The bench has been used to compare the chromatic performance of several coronagraphs, including the four-quadrant phase-mask (FQPM) coronagraph \citep{Bonafous2016DevelopmentAndCharacterization}, multi-stage four-quadrant phase-mask (MFQPM) coronagraph \citep{Galicher2014HighContrastImagingOn}, dual-zone phase-mask (DZPM) coronagraph \citep{Delorme2016LaboratoryValirdation}, vector vortex coronagraph (VVC), six-level phase-mask coronagraph \citep{Patru2018InlabTestingSixlevel} and the wrapped vortex coronagraph \citep[WVC,][]{Galicher2020AFamilyofPhaseMasks}.

One specific focus of the activities on the THD2 testbed has been the testing and comparison of WFS\&C techniques under equivalent conditions \citep{Potier2020ComparingFocalPlane, Herscovici-Schiller2018ExperimentalValidationNonlinear}, in particular pioneering the self-coherent camera \citep[SCC,][]{Mazoyer2014HighContrastImagingPoly, Mazoyer2013EstimationAndCorrection, Baudoz2012DarkHoleAndPlanet, Mas2012ExperimentalResults, Mas2010SelfCoherentCamera}. The SCC was the main estimation technique used since the inception of THD2, and was later complemented by the option to do PW probing \citep{Potier2020ComparingFocalPlane}.

In this section, we provide a general overview over the THD2 testbed. We describe the design and manufacturing choices for the optical masks that represent a Roman aperture and an HLC with similar performance. We also manufactured an SPC for THD2 but we are not considering it in this paper. We also briefly lay out the WFS\&C strategy deployed on THD2.

\subsection{Testbed setup}
\label{subsec:testbed_setup}

\begin{figure*}
   \resizebox{\hsize}{!}{\includegraphics{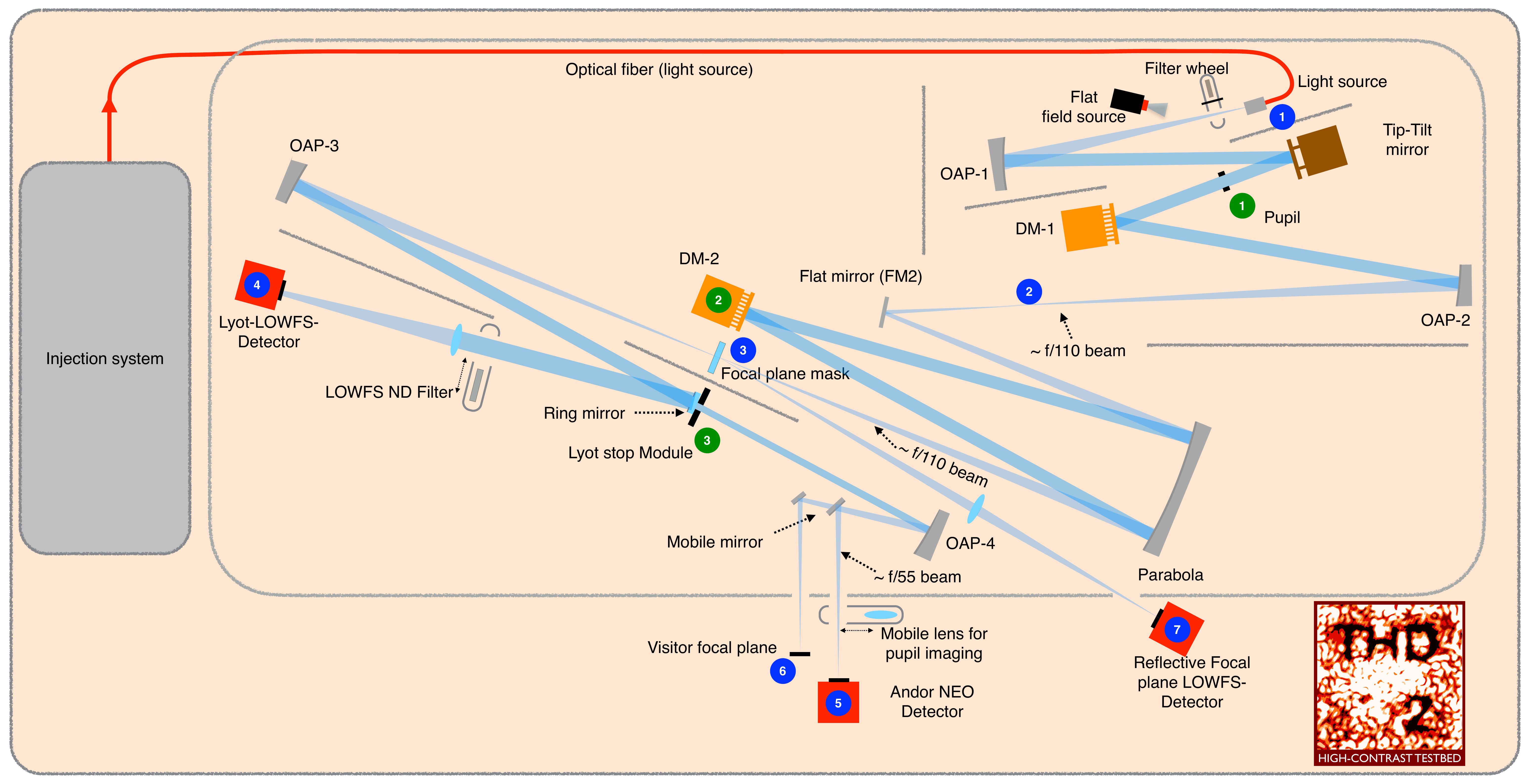}}
   \caption[THD2 testbed layout] 
   {\label{fig:thd2_testbed_layout} 
    THD2 testbed layout. Blue circle markers indicate focal planes and green circle markers indicate pupil planes. The laser injection system positioned on the left side injects the light through a fiber into the top right part of the testbed in this figure (blue circle 1). The available monochromatic laser diodes emit at wavelengths  638~nm, 705~nm and 783~nm, separately or combined. A filter wheel containing one clear position for normal imaging, a light trap for dark images and a flat mirror for injection of a flat field source is located right after the fiber. A tip-tilt mirror is located before the first pupil plane that contains the aperture mask (green circle 1). The beam then hits DM1, the out-of-pupil DM on THD2, converges through an unused focal plane (blue circle 2) and reaches the in-pupil DM2 (green circle 2). The following focal plane contains the focal-plane mask (blue circle 3) and then leads to the Lyot stop mount (green circle 3). A ring mirror around the Lyot stop reflects the light towards a LOWFS camera (blue circle 4), not used for the work in this paper. The reflection off of the HLC's central dot is imaged onto a second LOWFS camera (blue circle 7), used for closed-loop TT control in this paper. The transmitted light is focused on the science camera (blue circle 5) or alternatively directed to a secondary detector plane (blue circle 6) by means of an insertable flat mirror. A lens can be moved into the beam before the science camera to obtain pupil images, and a linear polarizer is located before this camera as well.}
   \end{figure*}

The THD2 testbed is housed in an ISO~7 pressurized cleanroom to minimize dust and contaminants. Temperature gradients are controlled within 0.1 degrees to reduce air turbulence, and several layers of enclosures stabilize the temperature around the beam path. A testbed layout is shown in Fig.~\ref{fig:thd2_testbed_layout}.
The light injection module uses lasers at wavelengths of 637~nm, 705~nm, and 785~nm, joined via dichroic fiber combiners. Short-term future work will replace these diodes with a controllable white light source. The module ensures precise injection of light into a monomode fiber, which acts as a point source for experiments. For details about the full testbed layout, we refer to the caption of Fig.~\ref{fig:thd2_testbed_layout}. The tip-tilt (TT) mirror has recently been changed to improve the correction of the residual turbulence and the vibration-induced TT on the bench. With a settling time of less than 0.7~ms, this new TT mirror should allow removing vibration effects up to $\sim$50~Hz. The pupil mask is held by a motorized mount that allows us to change between different pupil mask designs, one of them the Roman pupil, shown in Fig.~\ref{fig:pupil_masks_thd2}, left. The out-of-pupil DM1 is positioned at a distance of 27~cm after the first pupil plane. The FPMs are mounted on kinetic mounts, allowing for fast exchange of different coronagraph masks, among them two custom-made HLC masks. Two different LOWFS paths can be used depending on the coronagraph placed in the focal plane (blue circle 3 in Fig.~\ref{fig:thd2_testbed_layout}). One uses the light reflected from the area outward of the Lyot stop (LS) and directed towards a dedicated Lyot-LOWFS camera. This is used to stabilize the beam during closed-loop correction \citep{Singh2014LyotBasedLowOrder}, but was not used for the work in this paper. The second LOWFS uses the light reflected from the FPM and is focused on another LOWFS camera (blue circle 7 in Fig.~\ref{fig:thd2_testbed_layout}). This TT correction loop was used during all experiments in this paper. The LS assembly is motorized and can easily swap Lyot stops, like putting in the Roman LS, shown in Fig.~\ref{fig:pupil_masks_thd2}, right. The science camera is currently an Andor Neo camera using a fixed exposure time (10~ms), and variable entrance flux to simulate different observing configurations. The flux is measured in real time during the experiments using a splitting fiber that sends a small fraction of the flux (about 1\%) of the injection fiber into a calibrated photometer.

\subsection{Roman HLC on THD2}
\label{subsec:roman_hlc_on_thd2}

\subsubsection{Entrance pupil mask and Lyot stop}

\begin{figure}
   \resizebox{\hsize}{!}{\includegraphics{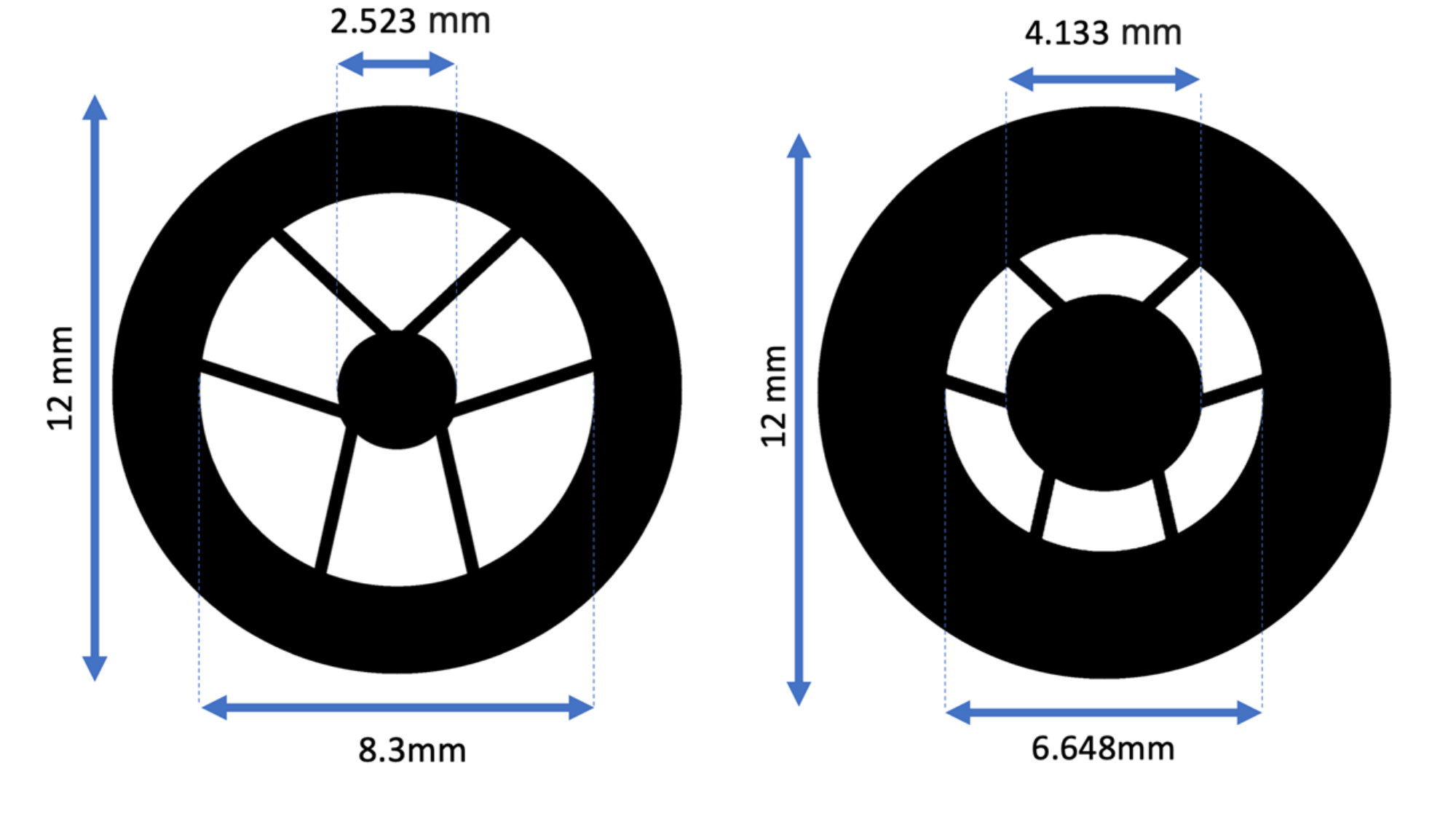}}
   \caption[Roman HLC THD2 masks] 
   {\label{fig:pupil_masks_thd2} 
    Entrance pupil mask (left) and Lyot stop mask for the Roman HLC configuration on THD2. The mask sizes have been scaled to the respective optical sizes on THD2, in particular to the 8.3~mm clear pupil on the in-pupil DM2. The LS diameter of 6.648~mm is 80\% of the pupil mask diameter. The strut width of the pupil mask and LS is 0.28~mm in both cases. The large central obscuration and non-radially symmetric supports are particularly challenging for coronagraphy.}
   \end{figure}

There are two main differences between Roman CGI and THD2 which impact the design of the coronagraphs. First, the type of DMs are different, which has an impact on the pupil size. Second, the first pupil plane (entrance pupil of the coronagraph) is reflective on CGI while transmissive on THD2. Due to the compactness of the Boston Micromachine DMs on THD2 (32$\times$32 actuators on a 1$\times$1~cm square DM surface for in-pupil DM2) and some known damaged actuators, the pupil size on the in-pupil DM is constrained to 8.3~mm. This sets the diameter of the central obscuration of a Roman-like pupil mask to 2.54~mm, and the width of the struts to 0.28~mm. A metal-cut pupil mask with the shape of the scaled Roman CGI entrance pupil has been produced for and installed on THD2, shown in Fig.~\ref{fig:pupil_masks_thd2}, left.
Both THD2 and Roman CGI have transmissive Lyot stops. The LS diameters are undersized compared to the entrance pupil. For the Roman LS on THD2, its outer radius is 6.629~mm and inner radius 4.150~mm. Equally produced from cut metal, it is shown in Fig.~\ref{fig:pupil_masks_thd2}, right.

\subsubsection{Simplification of the FPM design}

\begin{figure}
   \resizebox{\hsize}{!}{\includegraphics{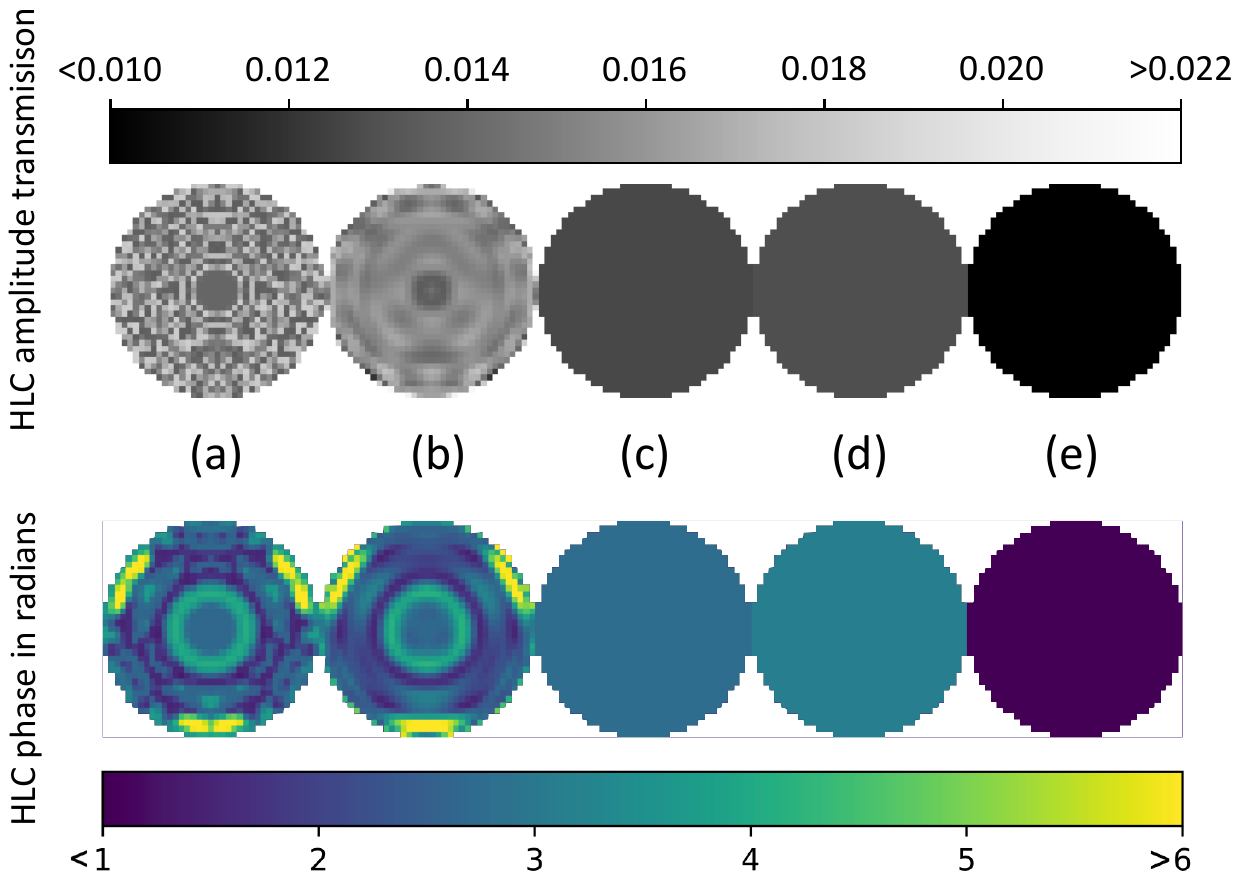}}
   \caption[HLC mask simplifications] 
   {\label{fig:fpm_simplifications} 
    The five different HLC mask designs explored for manufacturing on THD2. Option (a) is the original mask for Roman CGI \citep{Trauger2016HybridLyotCoronagraphFor}, (b) is its projection on 144 Zernikes, (c) is a projection on a single Zernike (piston), (d) is reduced to a $\pi$ phase shift and (e) is a classical Lyot coronagraph with an opaque mask of zero transmission and zero phase.}
   \end{figure}

\begin{figure}
   \resizebox{\hsize}{!}{\includegraphics{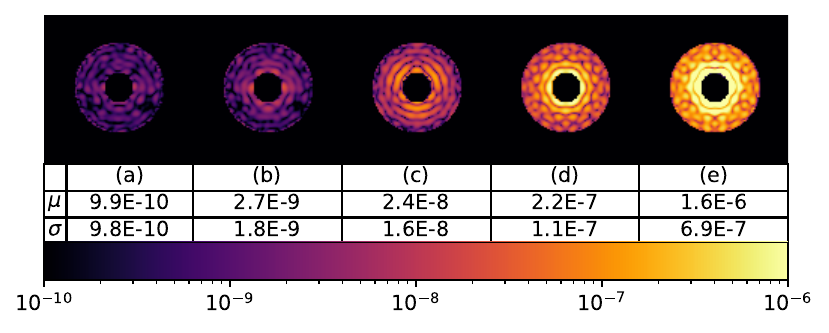}}
   \caption[DH solutions for FPM simplifications] 
   {\label{fig:dh_solutions_fpm_simplifications} 
    Simulated DH images (3--9~$\lambda/D$) for the five HLC mask solutions (a), (b), (c), (d), and (e) as described in the text and displayed in Fig.~\ref{fig:fpm_simplifications}. They all share the colorbar indicated at the bottom of the figure. The mean, $\mu$, and standard deviation, $\sigma$, of monochromatic contrast inside the DH are given in the table below each image. The imaging system was assumed aberration-free and no optimization of the DM shapes has been applied as we modify the mask design.}
   \end{figure}

For Roman, JPL has fabricated a custom HLC FPM described in Sec.~\ref{subsec:hybrid_lyot_coronagraph} and in \cite{Trauger2016HybridLyotCoronagraphFor}. This FPM is partially transmissive over a 5.6 $\lambda/D$ diameter with a transmission intensity of $2.45 \times 10^{-4}$ (1.57\% amplitude transmission) at the central wavelength of 575~nm. It also applies a $\simeq$2.9~rad phase shift at the central wavelength, with both transmission and phase being non-uniform, presenting a complex structure over the mask area \citep[see Fig.~3 in][]{Trauger2016HybridLyotCoronagraphFor}. For a custom HLC FPM on THD2, numerical simulations were conducted to assess whether the complex structure of this FPM could be simplified while maintaining performance. Five different mask configurations were tested: (a) the original phase and amplitude shapes proposed by JPL, (b) reduced phase and amplitude shapes projected onto 144 Zernikes, (c) a simplified version projected onto a single Zernike (piston) both for phase and amplitude, (d) an optimized amplitude mask with a $\pi$ phase shift, and (e) a classical Lyot mask with zero transmission and no phase change. The amplitude and phase of all five mask designs are shown in Fig.~\ref{fig:fpm_simplifications}.
Using monochromatic light at 575~nm (center of the band for the HLC on Roman), in absence of aberrations, simulations of circular DH images from 3 to 9 $\lambda/D$ were performed with a Roman CGI model and are displayed in Fig.~\ref{fig:dh_solutions_fpm_simplifications}.
The results indicated that while the more complex masks (a) and (b) provided better contrast of $9.8 \times 10^{-10}$ and $1.8 \times 10^{-9}$, even the simplified design (c) could achieve a mean DH contrast around $1.6 \times 10^{-8}$. Based on these simulation findings, we selected a simplified piston and phase mask design combined with optimized DM corrections for further testing, option (c). The other two solutions (d) and (e) were discarded due to insufficient performance.

\begin{figure}
   \resizebox{\hsize}{!}{\includegraphics{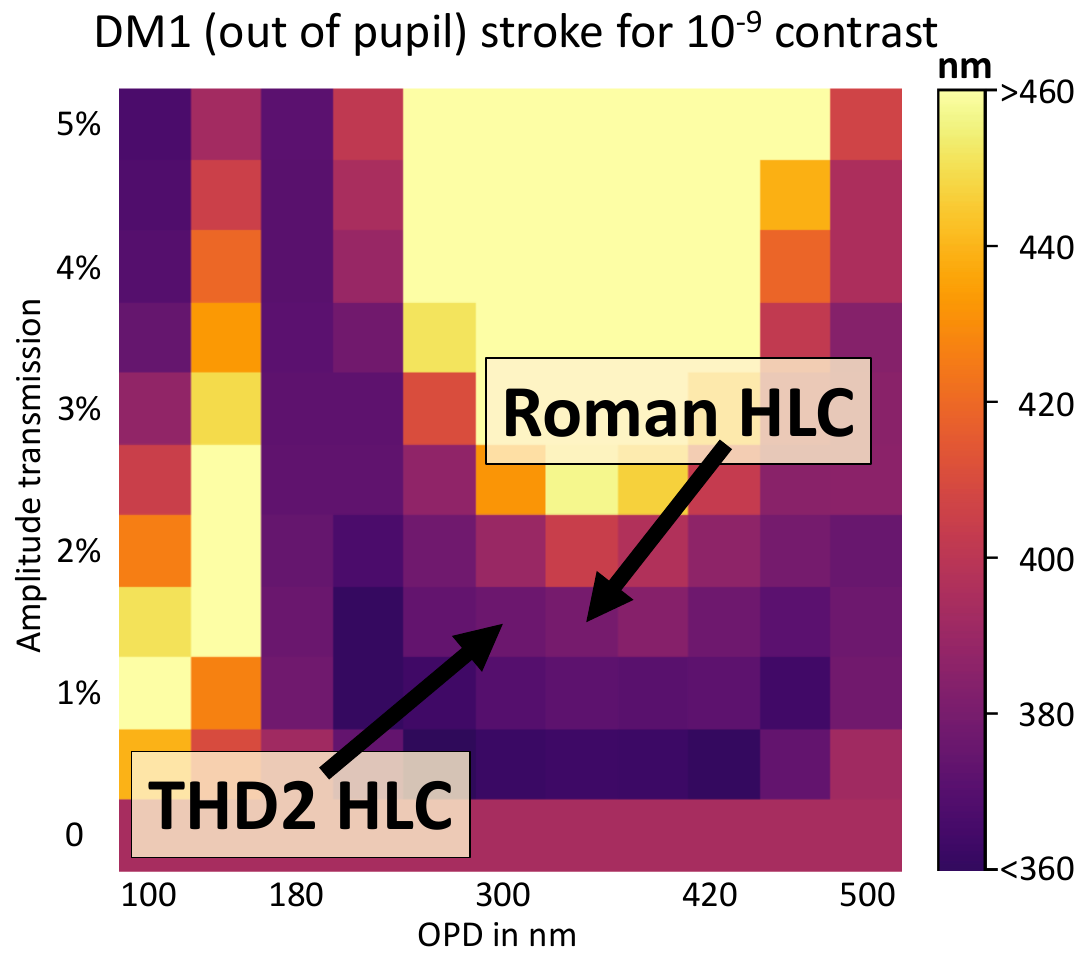}}
   \caption[DM stroke HLC FPMs] 
   {\label{fig:dm_stroke_hlc_fpms} 
    Required stroke of DM1 (out of pupil) on THD2 in nm to reach a DH contrast of $10^{-9}$ with an HLC mask designed with option (c) described in the text. The strokes are given as a function of amplitude transmission, on the y axis, and OPD in nm wavefront, on the x axis, of the HLC mask. This simulation assumes the two concrete DMs of THD2, and a correction at 700~nm, which is the worst case in terms of DM1 actuator stroke out of the three simulated wavelengths (700~nm, 740~nm, 780~nm). Using the same specs like Roman (1.57\% transmission and 2.87~rad phase) for an HLC on THD2 requires a DM1 stroke of 380~nm ptv and thus still leaves margin for aberration corrections. For easier manufacturing, the HLC for THD2 deploys a slightly lower OPD, still meeting specs. For details, see text.}
   \end{figure}

Further simulations with mask (c) were conducted to determine if a correction by the two DMs would be sufficient to reach a contrast of $10^{-9}$ in monochromatic light on THD2. Compared to Roman, THD2 has fewer actuators within the pupil (47 versus 28 across), and the Fresnel distance $L_F$ of the out-of-pupil DM actuator pitch is shorter ($L_F^{THD2}=0.43$ for a pitch of 0.3~mm and a central wavelength of 740~nm, compared to $L_F^{CGI}=1.71$ for a pitch of 0.99~mm and a central wavelength of 575~nm). The simulations used a Roman pupil with an 8.3~mm diameter, DM1 and DM2 with a 0.3~mm actuator pitch each, and a perfect field estimator in the focal plane. A DH was defined between 3 and 9 $\lambda/D$ for monochromatic corrections at wavelengths of 700~nm, 740~nm, and 780~nm, which are optimized with the anti-reflection coating of the protective windows of the DMs on THD2. Closed-loop simulations were performed with several HLC mask variants of the above-mentioned option (c), featuring different phase and amplitude levels. This showed that contrasts better than $10^{-9}$ could be achieved in closed-loop with a range of transmission levels (0--5\%) and optical path differences (OPD) between 100~nm and 500~nm of wavefront, assuming no testbed aberrations in the simulations. However, some solutions, particularly at $\lambda=700$~nm, required larger DM strokes. One limitation of THD2 is the maximum stroke of the out-of-pupil DM1, which is 480~nm peak-to-valley (ptv). Therefore, a mask design with a margin for realistic aberration correction with DM1 was necessary. Figure \ref{fig:dm_stroke_hlc_fpms} illustrates the DM1 stroke needed to achieve a contrast of $10^{-9}$ with the HLC mask designs that were used in simulation.
The results show an optimal parameter space for low DM1 stroke with transmission around $\sim$1\%, and an OPD between $\sim$250 and $\sim$450~nm ptv.
The original Roman design, with 1.57\% transmission and a phase shift of 2.87~rad (equivalent to 338~nm OPD at $\lambda=700$~nm), proves to be a solid option on THD2 too, requiring approximately 380~nm of stroke on DM1 and leaving sufficient stroke for aberration correction. Thus, the choice was made to fabricate an FPM with the same approximate transmission and phase specification like on Roman. Designs with lower transmission or smaller OPD offer even better performance, reducing the required stroke to under 360~nm.

\subsubsection{Design and fabrication of the HLC for THD2}

\begin{figure}
    \centering
    \includegraphics[scale=0.33]{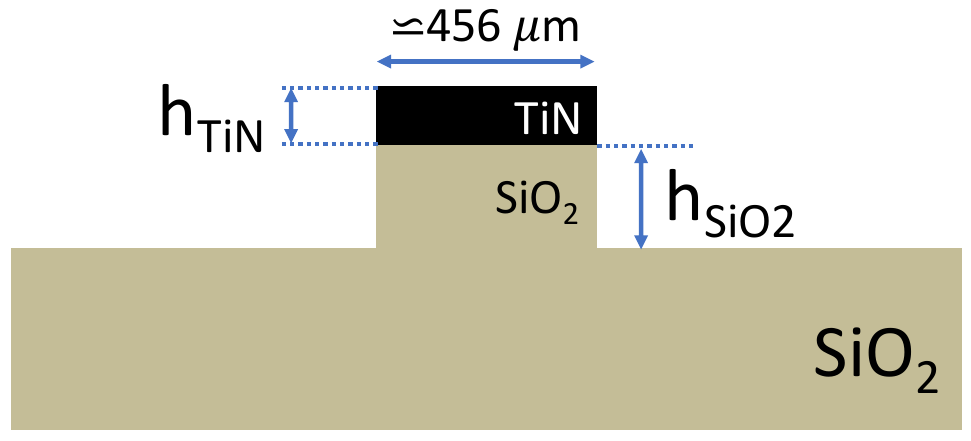}
   \caption[FPM design] 
   {\label{fig:fpm_design} 
    Sketch of the optical design of the HLC FPM as designed for the THD2 testbed. On top of a one-inch SiO$_2$ substrate, a SiO$_2$ step is used to optimize the OPD, and the TiN layer is deposited to create the required low transmission. The quoted diameter is for the longer wavelength range as specified in Tab.~\ref{tab:MaskDescription}. Sizes not to scale.}
   \end{figure}

The chosen FPM design for an HLC on THD2 contains a density spot deposited on a fused silica (SiO$_2$) spot to create the required OPD, see Fig.~\ref{fig:fpm_design}.
The absorptive material used in the top layer was titanium nitride (TiN). We performed transmission measurements on TiN samples with thicknesses from 100~nm to 250~nm \citep{Nicaise2022DevelopmentOfMicrowave} and derived the following empirical function giving the transmission, $T$, as a function of the TiN layer thickness, $h_{TiN}$, in nm:
\begin{equation}
    T = e^{-0.039~h_{\text{TiN}} - 0.354},
    \label{eq:tin_transmission}
\end{equation}
for a wavelength of 740~nm. Solving for the required material thickness given an expected transmission, we obtain:
\begin{equation}
    h_{\text{TiN}} = -9.1 - 25.64 \ln(T).
    \label{eq:height_requirement}
\end{equation}
For a theoretical mean transmission of 1.57\% ($2.45 \times 10^{-4}$ in intensity at 740~nm) this requires a TiN layer thickness of $h_{TiN}=204$~nm. Ellipsometry measurements on a TiN sample deposited in the same way gave us an optical index of 1.57 at 740~nm \citep{Nicaise2022DevelopmentOfMicrowave}.

Based on these measurements, specifications for the initial fabrication prototypes of a custom-made HLC were established. We kept the 1.57\% transmission for the mask in order to set the corresponding TiN layer thickness to 200~nm. This 200~nm TiN layer produces an OPD of 114~nm at a wavelength of 740~nm. To reach the full required OPD of 338~nm, a SiO$_2$ layer with a thickness of 493~nm would be needed below the TiN layer to complement it. However, to make the fabrication process easier, the SiO$_2$ thickness was reduced to 400~nm. This adjustment results in an OPD of 296~nm, which is, with $\sim$370~nm of stroke, still compatible with the DM1 requirements shown in Fig.~\ref{fig:dm_stroke_hlc_fpms}.

The mask's diameter must be 5.6~$\lambda/D$ at the central wavelength of 740~nm to match the prescription used on Roman. Using precise $f/D$ estimates on THD2 (focal length $f$ at the coronagraph plane: 913.4~mm $\pm$ 5~mm) and a pupil diameter of 8.3~mm, the required HLC mask diameter was determined to be 456 $\pm$ 3~$\mu$m. Finally, to allow us to work in two distinct wavelength bands, two lithographic masks were produced for THD2: One with a diameter of 456~$\mu$m (5.6 $\lambda/D$ at 740~nm, planned bandwidth: 700-780~nm) and one with a diameter of 416~$\mu$m (5.6 $\lambda/D$ at 675~nm, planned bandwidth 637-700~nm).

The HLC masks were manufactured on-site at Paris Observatory in the LUX laboratory (former GEPI) by first creating the fused silica step using standard 365~nm UV photolithography on a standard, one-inch SiO$_2$ substrate. Its height $h_{SiO_2}$ is defined by lapping the uncoated sector by ion reacting etching process. The second step is to deposit the TiN layer. Further optical and physical parameters, as well as the deposition process, of TiN layers can be found in~\cite{Nicaise2022DevelopmentOfMicrowave}. The main requirement is to ensure the alignment of the TiN deposition with the first etching which is essential to maintain the high-level performance. Dedicated structures have been designed on the edge of the photolithographic masks to ensure the reproducibility of the alignment. Layers of AR coatings were deposited afterwards limiting reflexion on each surface to less than 0.2\% between 600~nm and 800~nm. Final results of the manufactured components are shown in Table \ref{tab:MaskDescription}.

\begin{table}[h]
    \centering
    \begin{tabular}{|c|c|c|c|c|}  
        \hline  
         Bandwidth & \multicolumn{2}{c|}{700-780~nm}& \multicolumn{2}{c|}{640-700~nm}\\    
         & Specs& Measured &Specs& Measured\\
        \hline  
        Mask diam. ($\mu m$) & 456 & 455 & 416 & 414\\ 
        Align. error ($\mu m$) & 0 & <2 & 0 & <2 \\
        $h_{SiO_2}$ (nm) & 400 & 400 & 320 & 318\\
        $h_{TiN}$ (nm)& 200& 198& 270 & 271\\
        \hline  
    \end{tabular}
   \caption[FP Mask description]{\label{tab:MaskDescription} Comparison between manufactured and specified FPM to simulate a Roman HLC for the THD2 testbed.}

\end{table}

\subsection{WFS\&C setup on THD2}
\label{subsec:wfsc_setup_thd2}

The baseline WFS\&C setup for the creation of a DH on THD2 is the same as for Roman CGI, i.e. PW probing to estimate the electric field in the coronagraphic focal plane, and EFC to minimize the light in a pre-defined DH region. The linear optical model representing the THD2 testbed with the HLC is written in Python and contained in the public package Asterix\footnote{\url{https://github.com/johanmazoyer/Asterix}} on GitHub. It is used to create the observation and control matrices which are then passed on to the current Labview testbed controls to close the loop on hardware. This WFS\&C loop runs very efficiently at about 3--5~Hz in monochromatic light. A control software upgrade is currently ongoing to migrate to the C++/Python-based controls provided by the catkit2 package \citep{Por2024catkit2}, which will mitigate risks related to old control infrastructure components and increase operational flexibility at no loss of loop speed.


\section{Probe choices for optimal pairwise estimation}
\label{sec:probe_choice}

Efficient probes are essential for PW estimation to modulate the speckle intensity and recover the electric field in focal-plane wavefront sensing. Originally, \cite{Give'on2011PairwiseDeformableMirror} introduced sinc-based probes in the pupil plane, designed to create a spatially uniform electric field across rectangular regions of the science image. This approach ensured that every pixel in the DH was modulated by all probes in a set.

Subsequent work, such as that by \cite{Groff2016MethodsLimitationsFocal}, explored the relationship between probe amplitude and estimation performance, considering also the effects of errors on the assumed probe shape. However, these analyses were limited to amplitude variations and assumed fixed probe shapes, leaving the impact of alternative probe designs largely unexplored.

In this section, we examine three distinct shapes of probes used in our analyses. First, we revisit the ``classic'' sinc-sinc-sine probes, adapted for use with the Roman-like HLC on the THD2 testbed. Second, we describe single-actuator probes, as proposed by \cite{Potier2020ComparingFocalPlane}, which have been the default approach for PW probing on THD2 since their implementation. Finally, we introduce ``sharp'' sinc probes, inspired by the compact nature of single-actuator probes, aiming to create probes even narrower than the actuator influence function. This exploration offers a broader understanding of how probe design impacts PW estimation.

\subsection{Sinc-sinc-sine probes}
\label{subsec:sinc_sinc_sine_probes}

The original DM probes used for PW probing as introduced by \cite{Give'on2011PairwiseDeformableMirror} are based on the Fourier relationships between the pupil plane and the focal plane. The goal was to identify DM commands in the pupil plane that would modulate a symmetric DH area in the scientific focal plane as uniformly as possible. This reasoning behind probe choice has been used for over a decade for ground-based systems on testbeds \citep{Thomas2010LaboratoryTestOfApplication} and on-sky \citep{Matthews2017ElectricFieldConjugation,Fusco2016SAXOtheSPHERE,Ruffio2014NonCommonPathAberrations}, and the strategy has been adopted as such for demonstrations of space-based systems like in operations at HCIT \citep{Noyes2023TheDecadalSurveyTestbedTwo,Ruane2022BroadbandVectorVortex}. The sinc-sinc-sine probes will also be the baseline probe shape for Roman CGI \citep{Cady2025HighOrderWavefront}.

A simple rectangle in the focal plane can be constructed by convolving a top-hat function across the x axis and the y axis of the focal plane each. A further convolution with a symmetric pair of delta functions positions an equal duplicate of the rectangle to either side of the image plane, covering the anticipated DH area, as exemplified in Fig.~1 by \cite{Give'on2011PairwiseDeformableMirror}. The Fourier transform of the top-hat functions is the multiplication of two cardinal sine functions (sinc) across either axis of the DM. The convolution with delta functions transforms into a multiplication by a sine or cosine along one axis. We can write the associated probe phase on the DM, $\Psi_j (u,v)$, as:
    \begin{equation}
    \Psi_j(u,v) = \sinc(\xi_1 u) \sinc(\xi_2 v) \sin(2\pi \xi_3 u+\theta_j),
    \label{eq:classic_sinc_probes}
    \end{equation}
where $\xi_1$ and $\xi_2$ are the width and height of the created rectangles in the $u$ and $v$ direction, respectively, $\xi_3$ is the center position of the rectangles, all in units of spatial frequency, here $\lambda/D$, and $\theta$ is the spatial phase shift of the probe. During PW probing, the goal is to approximate the phase in Eq.~\ref{eq:classic_sinc_probes} as best as possible with the DM probe commands $\bm{\psi}_j$.
Since two rectangles placed next to each other in the focal plane will leave a less well modulated strip along one axis, at least one probe pair of a full probe set should invert the pupil plane coordinates $\bm{u}$ and $\bm{v}$ in Eq.~\ref{eq:classic_sinc_probes} to obtain a rectangle pair rotated by 90 degrees for full coverage.

To create appropriate sinc probes for the HLC setup on THD2, we used a set of three probe pairs as defined by Eq.~\ref{eq:classic_sinc_probes}, sampled on the 32 $\times$ 32 actuator grid of the in-pupil DM2, of which 28 lie within the pupil mask. For a DH spanning 3--9 $\lambda/D$, we used $\xi_1=12~\lambda/D$, $\xi_2=24~\lambda/D$ and $\xi_3=6~\lambda/D$. To create a set of three probes, we applied phase shifts, $\theta$, of $1/4~\pi$, 0 and $3/4~\pi$. Finally, since the Roman HLC setup works with an extremely obscured pupil, the probes have been shifted in the DM plane to one of the clearest effective areas of the Lyot stop, which is the same strategy as followed by \cite{Matthews2017ElectricFieldConjugation} and Roman CGI. Two of the three resulting sinc-sinc-sine probes used with the HLC on THD2 are shown in the top row of Fig.~\ref{fig:probes_in_pupil}.

\subsection{Single-actuator probes}
\label{subsec:single_actuator_probes}

The single-actuator probes have first been formally introduced by \cite{Potier2020ComparingFocalPlane} by a motivation to avoid mechanical constraints and non-linear coupling in the effectively produced surface by the DM, given a command.
The prerequisite for this is a well-constrained actuator influence function in the optical model, which for THD2 was provided by \cite{Mazoyer2014DeformableMirrorInterferometric}. Subsequently, due to their straightforward implementation and no known disadvantage with respect to classic sinc probes, they have been used on SCExAO \citep{Ahn2021SCExAOTestbedDeveloping}, MagAO-X \citep{Haffert2023ImplicitElectric,Lieberman2024ANalyzingMisalignment}, and the HiCAT testbed, among others. On HiCAT, they were also compared to probes obtained through an inverse propagation using the optical model's influence matrix \citep{Will2021WavefrontControlAlgorithmic}, but no systematic contrast performance difference was found \citep{Pourcelot2023LowOrderWavefront}.

Following \cite{Potier2020ComparingFocalPlane} and \cite{Ahn2021SCExAOTestbedDeveloping}, we identified the most impactful DM2 actuators by poking all of them individually in simulation, by an equal amount, and measuring the resulting normalized change in DH mean contrast.
This simulated test showed us that all actuators that are completely unobscured by the pupil mask and Lyot stop are good candidates to be used as PW probes. To enable a fair comparison with the classic sinc probes produced in Sec.~\ref{subsec:sinc_sinc_sine_probes}, we chose as the main actuator the same like at the strongest actuator in the sinc probes, actuator 298. Finally, following \cite{Potier2020ComparingFocalPlane} to ensure the best possible coverage in the DH, we chose two more actuators adjacent to the initial one, and displaced along either DM axis, to form a set of three: actuators 297, 298 and 266. The location of actuator 298 in the THD2/HLC pupil is shown in the bottom left panel of Fig.~\ref{fig:probes_in_pupil}.

\subsection{Sharp sinc probes}
\label{{subsec:sharp_sinc_probes}}

One of the main motivations for the single-actuator probes is that they are narrower on the DM than sinc-sinc-sine probes, which makes their impact in the focal plane wider. Since the DM influence function can be approximated with a Gaussian, and the Fourier transform of a Gaussian is a Gaussian again, a single-actuator probe leads to a wider modulation, with a non-uniform estimation gain across the inner and outer pixels in the DH. A different influence function will translate to a different estimation gain between the inner and outer pixels in the DH.
This means that the narrower the DM probe command, the more uniform the modulation and thus electric field estimation in the focal plane.

The original sinc-sinc-sine probes aim to achieve this uniform modulation across a designated DH region. However, due to limitations of the DM, additional steps are necessary, such as convolving the probes with a Dirac delta function and rotating rectangular probe pairs to achieve full DH coverage, as described in Sec.~\ref{subsec:sinc_sinc_sine_probes}. As an alternative, we propose ``sharp sinc'' probes that simply use sinc functions along each axis, without a sine:
    \begin{equation}
    \Psi_j(u,v) = \sinc(\xi_1 u) \sinc(\xi_2 v),
    \label{eq:sharp_sinc_probes}
    \end{equation}
with $\xi_1 = \xi_2$ for all probes in a set.
These probes are made narrower by increasing their spatial frequency, allowing them to cover the entire DH and beyond. This approach involves undersampling the sinc functions, which is acceptable in this case: the undersampled representation of the sinc functions still evokes the desired Fourier response in the focal plane. It creates some high-spatial frequency aliasing well beyond the considered DH which does not impact the estimation.

\begin{figure}
   \resizebox{\hsize}{!}{\includegraphics{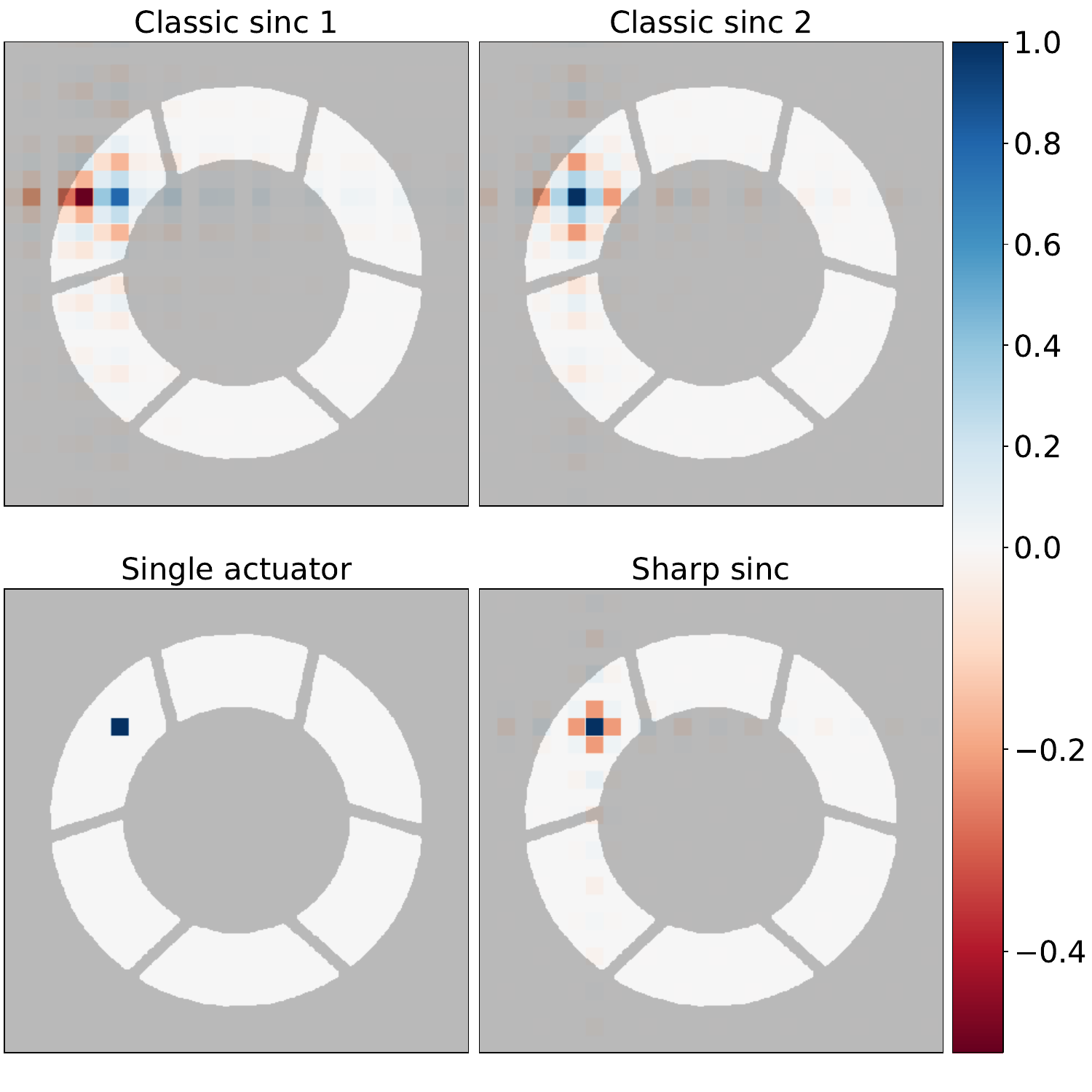}}
   \caption[Probes in pupil] 
   {\label{fig:probes_in_pupil} 
    DM commands of the three probe shapes used for the electric field estimation, overlaid on the Roman LS of the HLC configuration of THD2. All are normalized to a unit amplitude of one. The probes are applied on the in-pupil DM2 in all cases. \textbf{Top:} Two of the three probes in the sinc-sinc-sine set. The third one is equal to the top right one but rotated by 90 degrees, located in the same place on the DM. \textbf{Bottom:} One of the single actuators (left) and sharp sinc probes (right). Both are centered on the same actuator. The other two probe of the set are a copy of the ones shown here, respectively, with a one-actuator offset to the left, and to the top.}
   \end{figure}

Undersampling provides the added benefit of ensuring that adjacent DM actuators move in opposite directions (one pushed, the other pulled). This creates a probe command even narrower than the single-actuator influence function, resulting in a broader response in the focal plane, shaped like a square. While this approach might be constrained in practice by inter-actuator effects that limit the extent of ``stretching'' \citep{Blain2009NonAdditiveFunctionMEMS_DM}, it can still provide valuable insights into how probe width affects estimation performance. For the sharp sinc probes on THD2, we used $\xi_1= \xi_2 = 48~\lambda/D$. Like for the single-actuator probes, the first of the sharp sinc probes is centered on actuator 298, with the other two spatially phase shifted to the adjacent actuators 297 and 266. One of the three sharp sinc probes, centered on actuator 298, is shown in the bottom right panel of Fig.~\ref{fig:probes_in_pupil}.

\subsection{Pair-wise modulation of the DH, and nominal contrast performance}
\label{subsec:pw_estimation_accuracy}

\begin{figure}
   \resizebox{\hsize}{!}{\includegraphics{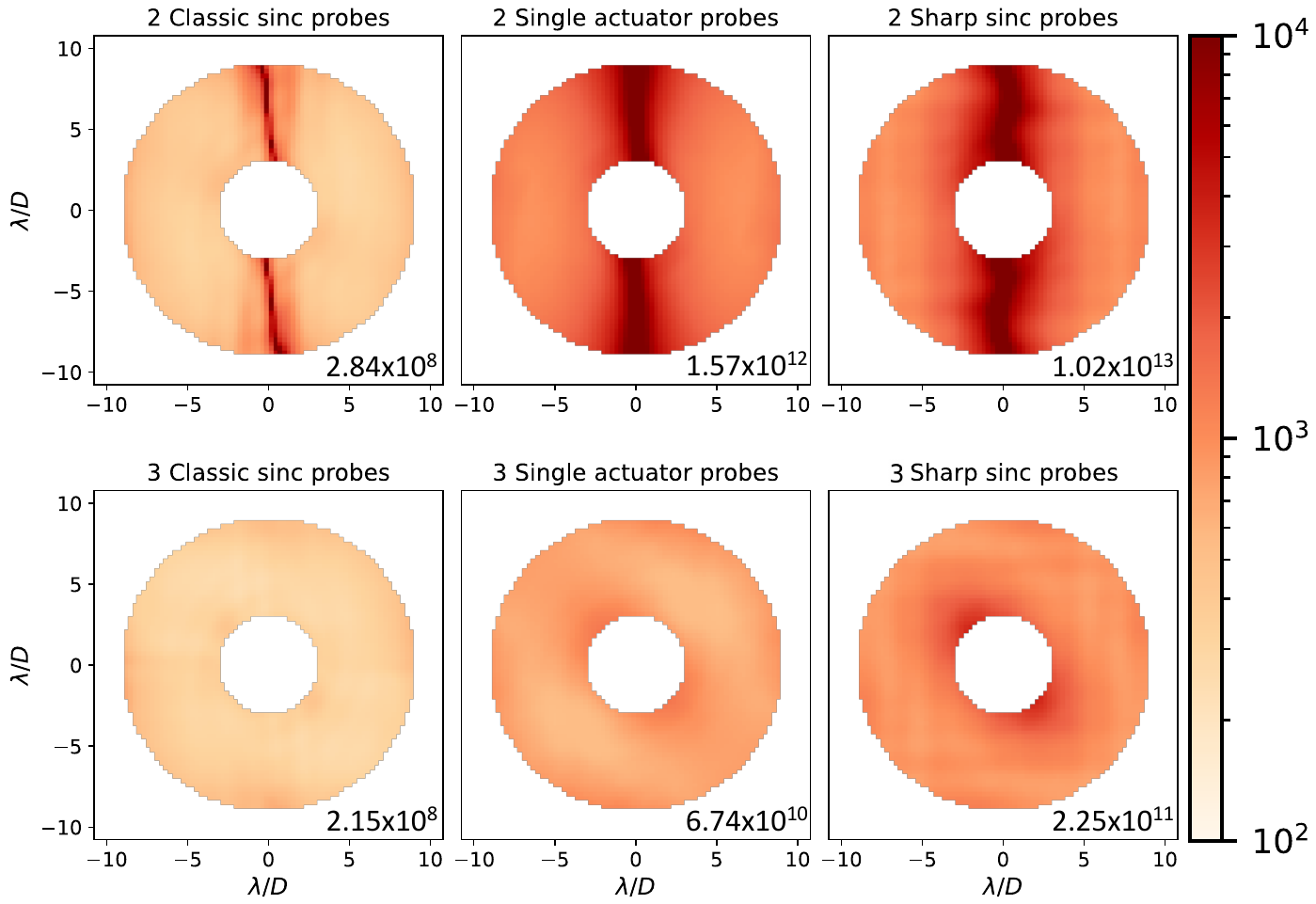}}
   \caption[DH detection map] 
   {\label{fig:DH_detection_map} 
    Inverse of the singular values of the pseudo inverse of the observation matrix, $H$, at each pixel of the DH (3--9~$\lambda/D$), for all three sets of probe shapes described in Sec.~\ref{sec:probe_choice}. High values represent poorly estimated pixels. All maps are plotted on the same scale. The condition numbers for the respective matrices are noted in their respective panel. \textbf{Top:} Modulation maps for sets of 2 probe pairs in a set, for all three probe shapes. All of them leave a strip of poorly modulated pixels along the central detector axis.
    \textbf{Bottom:} Modulation maps for sets of 3 probe pairs in a set, for all three probe shapes.
    In all three cases, using three probe pairs renders these maps very uniform which fulfills the necessary condition in Eq.~\ref{eq:necessary-condition_b} and qualifies the probes for efficient electric field estimation. }
   \end{figure}

To determine the effectiveness of all three probe sets described in Sec.~\ref{sec:probe_choice}, we first look at the modulation coverage of the pixels in the DH. Following \cite{Potier2020ComparingFocalPlane}, we calculate the inverse of the singular values of the pseudo inverse of the observation matrix, $H$, at each pixel of the science detector, and we display these maps in Fig.~\ref{fig:DH_detection_map} for all three shapes of probes described previously, and for a set of two and three probe pairs.
The displayed focal-plane maps indicate poorly modulated pixels with high values and well estimated pixels with low values. Formally, the ratio between the highest and lowest eigenvalue of the observation matrix, $H$, is expressed by its condition number, quoted in each of the panels in Fig.~\ref{fig:DH_detection_map}. If only two probe pairs were used, all three probe shapes leave an unmodulated strip of pixels along the central axis. In this case, the single-actuator and sharp sinc probes modulate the pixels the worst, which is reflected in their significantly higher condition number. \cite{Potier2020ComparingFocalPlane} already showed that it is important to pick actuators close to each other to avoid additional periodic patterns of poorly modulated pixels in the case of single-actuator probes. By including a third probe that breaks the symmetry along the central axis in each set, the modulation across the DH is much more uniform (Fig.~\ref{fig:DH_detection_map}, bottom row). This is also reflected in the lowering of the condition number for single actuator and sharp sinc probes, each by two orders of magnitude, when including a third probe pair in the set. The condition number of the classic sinc probes improves only marginally with a third probe pair.
While the three probe shapes have a slightly different global offset in their maps when three probe pairs are used for each, they are all very uniform. This uniformity fulfills the condition in Eq.~\ref{eq:necessary-condition_b} in all pixels of the DH, thus making these probe sets sufficient for electric field estimation.

With the maps from Fig.~\ref{fig:DH_detection_map} suggesting that all pixels in the focal plane will be sufficiently modulated for an electric field estimation with all three probe shapes, we run a WFS\&C experiment both in simulation and on hardware to assess the closed-loop performance of the probes. We use PW estimation with all three probe shapes described in Sec.~\ref{sec:probe_choice}, and EFC for wavefront control. For all WFS\&C runs in this paper, both in simulation and on hardware, the control matrix regularization parameter was held constant in all cases, and throughout all iterations. This is not optimal for best results on the testbed, however, this conservative approach allowed us to focus on a comparison of estimation effectiveness. Equally, the loop gain was held constant at 0.8 in simulation and 0.4 on hardware throughout all runs presented in this section. To provide the same level of DH intensity modulation between all probe shapes, we scale their amplitudes such that all of them add a $10^{-6}$ mean DH contrast when applied to the DM, across all iterations of the loop.
This was an arbitrary initial choice since probes of $10^{-6}$ added intensity have shown to perform well under ``standard'' conditions - the modulation is visible by eye in the probed images, meaning they are strong enough, and the loop converges, meaning they are not too strong.

All loops, in simulation and on hardware, were run in a monochromatic wavelength of $\lambda=783$~nm. In the noiseless simulations, we inject a random phase aberration with a power spectral density proportional to the power 3 of the spatial frequency, with 20~nm root-mean-square (rms) across the pupil. We also injected a 10\% rms amplitude aberrations, with a power spectral density proportional to the inverse of the spatial frequency. On hardware, we started the runs in Fig.~\ref{fig:contrast_iterations_hw} from pre-corrected DM settings, making the loops start at a lower contrast compared to the simulations. The flux was decreased on purpose to start from this less aberrated level without saturating the focal plane. In this configuration, the testbed camera does not have the dynamical range necessary for a correction below the level shown here. The main point here is to provide equal grounds to compare the three different probe shapes in a representative WFS\&C loop.

\begin{figure}
   \resizebox{\hsize}{!}{\includegraphics{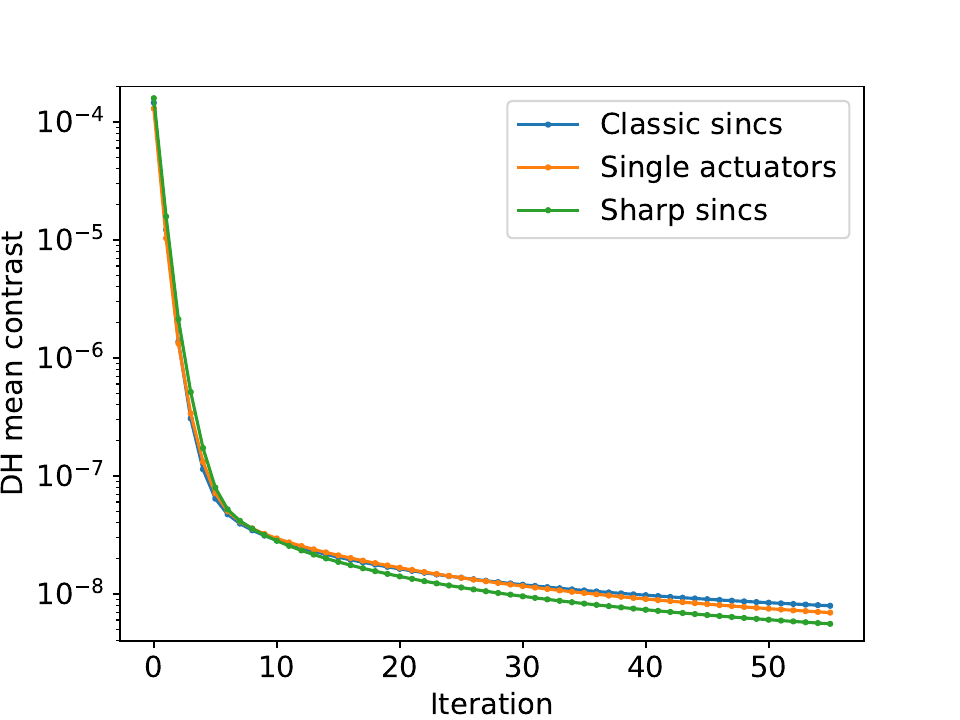}}
   \caption[Contrast iterations in simulation] 
   {\label{fig:contrast_iterations_sim} 
    Contrast iterations in \textbf{simulation}, using EFC to calculate DM commands. The data sets were taken sequentially with single actuators (blue), classic sinc probes (orange) and sharp sinc probes (green), all using 3 pairs in the PW estimation. The EFC regularization was held constant across all iterations. The loop was run at a wavelength of $\lambda=783$~nm and with a loop gain of 0.8, and the mean contrast was measured in a circular DH from 3--9~$\lambda/D$. All three shapes of probes were applied with a probe amplitude that adds a mean contrast of $10^{-6}$ to the DH, in each iteration. The single-actuator probes provide a slightly better contrast than the classic sinc probes at higher iteration numbers (after iteration $\sim$45), while the sharp sinc probes start performing better sooner (after iteration $\sim$15). The loop does not converge in the data shown here, but it reaches a level below $10^{-8}$, which is lower than the comparison hardware data shown in Fig.~\ref{fig:contrast_iterations_hw}. The simulations are noiseless.}
   \end{figure}

\begin{figure}
   \resizebox{\hsize}{!}{\includegraphics{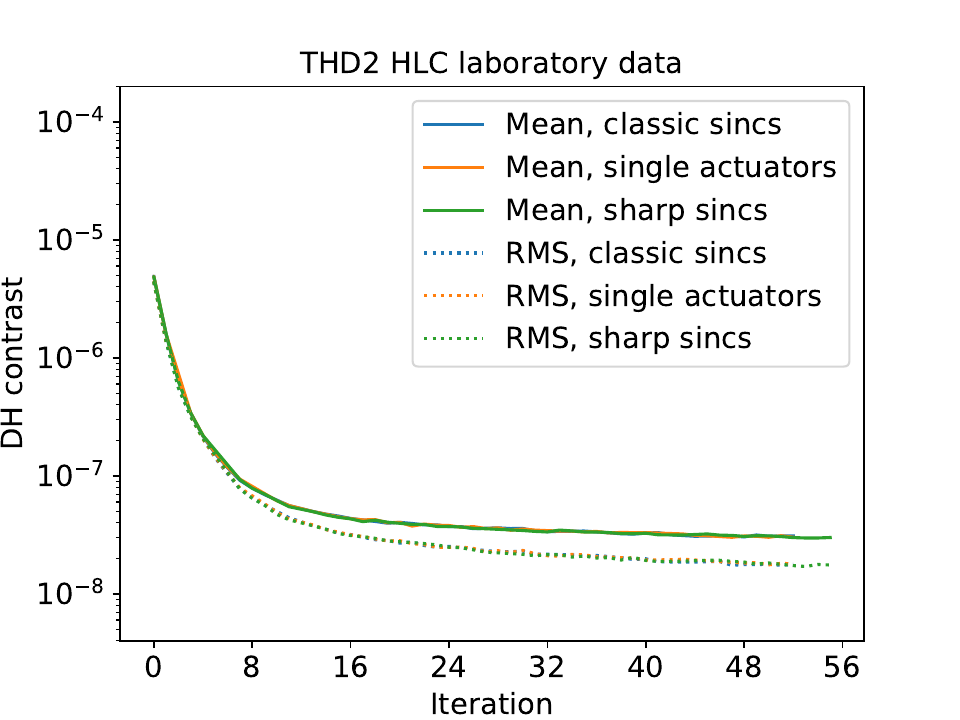}}
   \caption[Contrast iterations on hardware] 
   {\label{fig:contrast_iterations_hw} 
    Contrast iterations on \textbf{hardware}, using EFC to calculate DM commands. The data sets were taken sequentially with single actuators (blue), classic sinc probes (orange) and sharp sinc probes (green), all using 3 pairs in the PW estimation. The loop was run at a wavelength of $\lambda=783$~nm and with a loop gain of 0.4, and both the mean (full lines) and rms (dotted lines) contrast was measured in a circular DH from 3--9~$\lambda/D$. All three shapes of probes were applied with a probe amplitude that adds a mean contrast of $10^{-6}$ to the DH in each iteration. The loops were started from the same pre-corrected DM settings each time, at a flux level that would not saturate the focal plane. This limited the achieved contrast later on slightly, but the bench is limited by diffusion at this level anyway. All three probes reach a DH mean contrast of $3.0\times10^{-8}$ and a DH rms contrast of $1.7\times10{-8}$ after 50 iterations, showing no performance difference at moderate probe amplitudes.
    }
   \end{figure}

Once we close the DH loop, we measure the contrast inside a 3--9~$\lambda/D$ DH at each iteration of correction. The resulting contrast convergence plots are displayed in Fig.~\ref{fig:contrast_iterations_sim} for the simulated case and in Fig.~\ref{fig:contrast_iterations_hw} for the hardware experiment on THD2.
The best DH mean contrast on hardware with the HLC on THD2 is $3\times10^{-8}$ after 50 iterations. This is not quite as deep as the DH mean contrast achieved in simulation. There, the first 15 iterations perform very similar like on hardware, with minimal difference between the three probe shapes. After this point, the sharp sinc probes provide the steepest contrast improvement, followed by the single actuators. However, the contrast difference after 50 iterations in simulation remains marginal: $7.8\times10^{-9}$ for classic sinc probes, $6.8\times10^{-9}$ for single-actuator probes and $5.4\times10^{-9}$ for sharp sinc probes.

The most likely reasons for the performance difference between simulation and hardware is the known scattered light on the testbed, as well as residual errors in the testbed alignment, for example between the pupil mask and the Lyot stop, which has very little margin. One way to separate out the scattered light caused by diffusion that appears in the mean and does not exist in simulation is to measure the rms contrast (dotted lines) in the DH instead of the mean value (full lines). This captures only the level of speckle noise in the images which limits the detection of point sources in the DH. The rms contrast level reached by all three probe types on hardware is $1.7\times10^{-8}$.

In both cases, simulation and hardware, the contrast progression follows the same pattern with a sharp decrease initially in the first $\sim$10 iterations, and a shallower decline afterward probably explained by the limited number of EFC modes kept in the control matrix inversion. With the different loop gain between hardware and simulations, we constitute that its initial convergence is the same: while the simulation run reaches a contrast rms of $2\times10^{-8}$ after $\sim$15 iterations (assuming mean and rms contrast being the same in simulations due to lack of incoherent light), the hardware run reaches the same level in $\sim$40 iterations at half the loop gain. Another contribution to the remaining discrepancy between hardware and simulation results is that the optical model is not perfect and once the contrast is low enough, the remainder of the loop is limited by the testbed model errors. 

\begin{figure}
   \resizebox{\hsize}{!}{\includegraphics{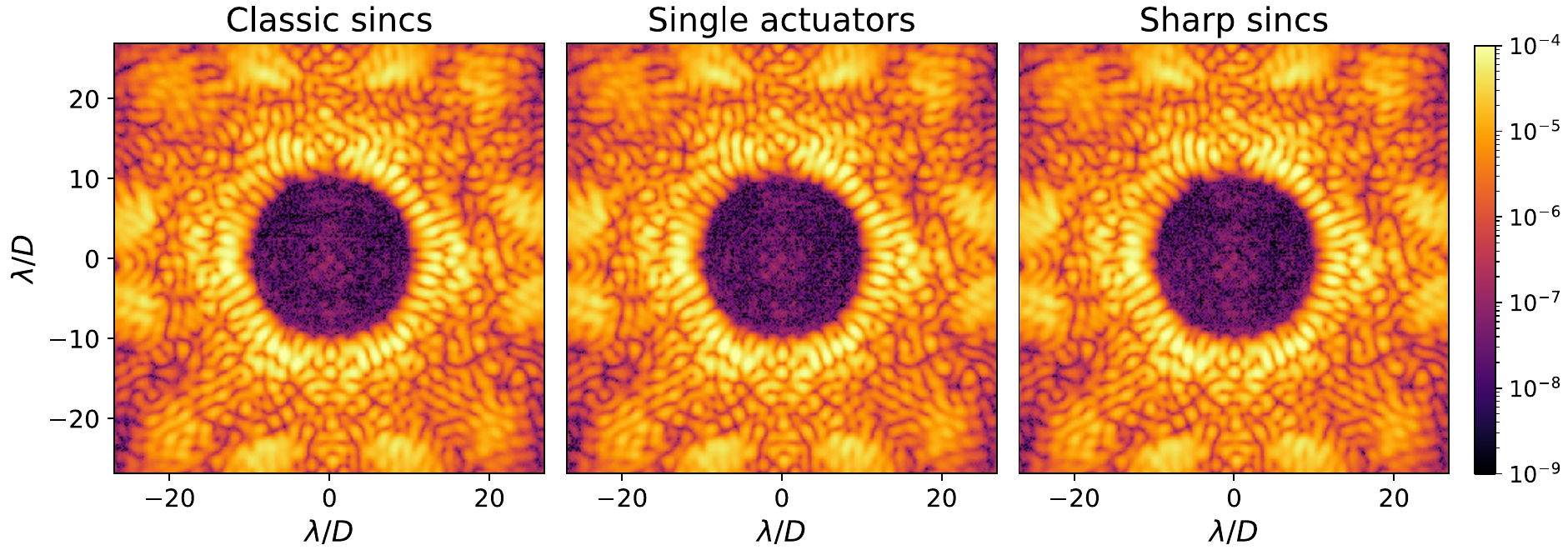}}
   \caption[Dark holes on hardware] 
   {\label{fig:dark_holes_hw} 
    DH images towards the end (iteration 50) of all three closed-loop WFS\&C runs shown in in Fig.~\ref{fig:contrast_iterations_hw}. The residual speckle pattern and intensity is very similar in all three cases, confirming the same level of contrast performance between probe shapes in this case. All three shapes of probes, sinc-sinc-sine, single actuators and sharp sincs, were applied with a probe amplitude that adds a mean contrast of $10^{-6}$ to the DH in each iteration. We conclude that there is no difference in contrast performance between the three probe shapes at moderate probe amplitudes.}
   \end{figure}

Figure \ref{fig:dark_holes_hw} shows the images taken during the hardware experiments at iteration 50 of all three contrast convergence plots shown in Fig.~\ref{fig:contrast_iterations_hw}. 
Between the three loops with different probe shapes, the speckle pattern at the end of the loop is the same, their difference revealing only white noise. The contrast performance in a representative DH loop with moderate-amplitude probes is thus the same for any of the three tested probe shapes, both in numerical simulations and on hardware.


\section{Non-linearities and performance trade-offs in high-amplitude pairwise probing}
\label{sec:probe_linearity}

The moderate-level probe amplitude of $10^{-6}$ was initially an arbitrary choice, as it was both strong enough to enable a sufficient signal-to-noise ratio (SNR) for the estimation and faint enough for all probes to justify the linear expansion of the electric field. However, there is a high interest to use higher-amplitude probes for an even higher SNR in the probe images. This would allow missions like Roman CGI to either run estimations on faint targets, and thus eliminate slews to higher-flux calibration stars, or calibrate a DH loop with shorter exposure times. In either case, higher-amplitude probes are highly desirable as they would liberate time for science observations

In this section, we investigate the contrast performance of high-amplitude probes on hardware, compare the linearity regimes of classic sincs and single actuators for the purpose of PW estimation, and present a way to exploit high-amplitude probes for fast estimation at low flux.

\subsection{Closed-loop performance of high-amplitude probes on hardware}

In Sec.~\ref{subsec:pw_estimation_accuracy}, we observed that for moderate probe amplitudes, all probe shapes perform equally well in a closed-loop system. This indicates that scaling all three used probe shapes to an amplitude equivalent of $10^{-6}$ relative mean contrast in the DH keeps them within a range that allows the linear approximation of the electric field in the PW formalism in Eq.~\ref{eq:E_im}.

\begin{figure}
   \resizebox{\hsize}{!}{\includegraphics{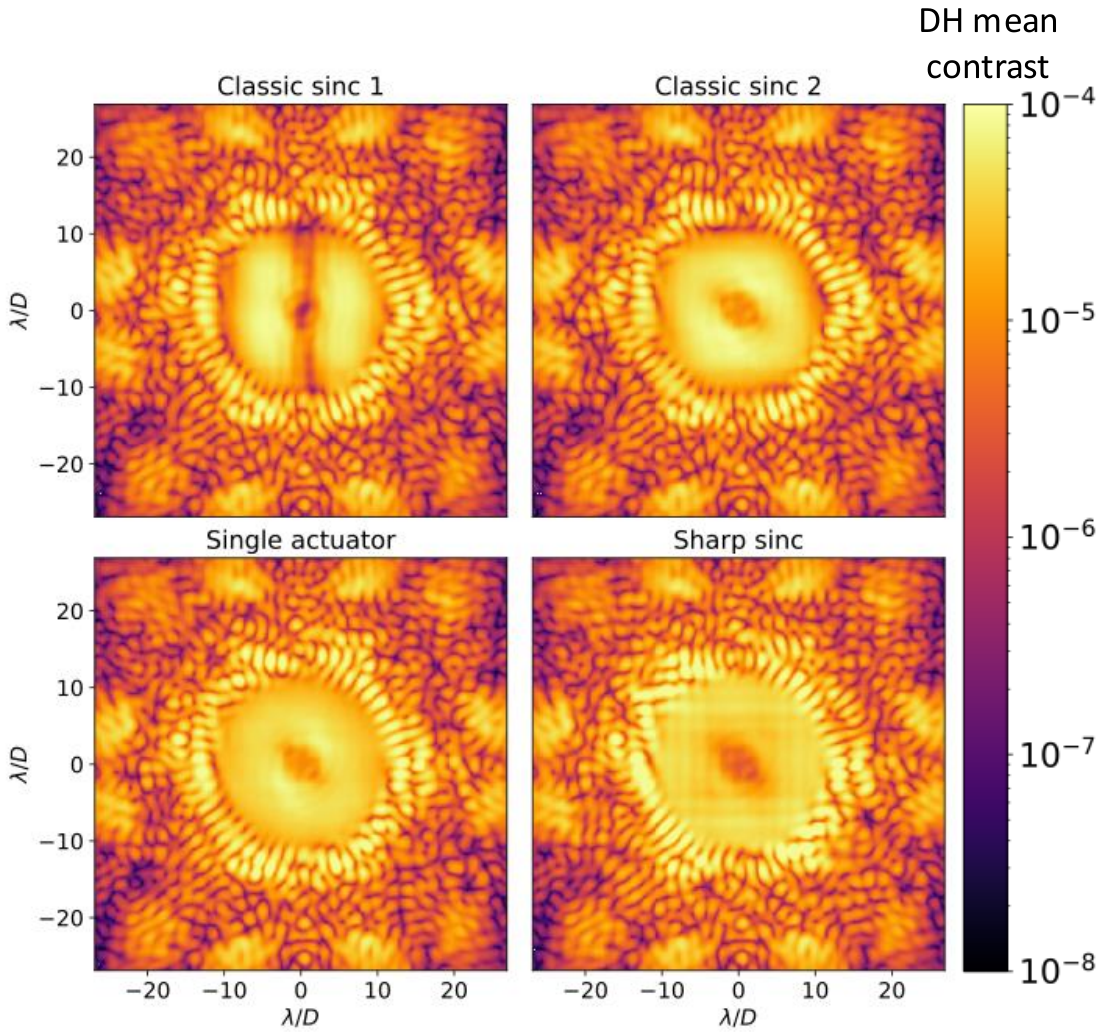}}
   \caption[Probe images on hardware] 
   {\label{fig:probe_images} 
    Probe images from experimental measurements showing the intensity distribution in the focal plane for different probe shapes at an added intensity of $36\times10^{-6}$ in the DH. \textbf{Top:} Two of the three sinc probes as described in Sec.~\ref{subsec:sinc_sinc_sine_probes}, and shown in Fig.~\ref{fig:probes_in_pupil}. \textbf{Bottom:} Single-actuator probe centered on actuator 297 (left) and a sharp sinc probe (right). The stripes in the bottom right panel originate from the sharp sinc probe itself, it is aliasing that is introduced by undersampling the sinc with respect to the highest Nyquist-sampled spatial frequency on the DM (see Eq.~\ref{eq:sharp_sinc_probes}).
    These very strong probes provide here an SNR of $\sim$480 and would thus allow to create a DH on a fainter target, or with shorter exposure times.}
   \end{figure}

To identify the upper limit on probe amplitude within the linear approximation, we scaled the original probe amplitudes by factors 1--7 and ran a series of WFS\&C loops on hardware under identical conditions for all three probe shapes. Images of the stronger probes, scaled to 36 times the $10^{-6}$ mean contrast (six times stronger in amplitude than the baseline case), are shown in Fig.~\ref{fig:probe_images}.
With a very high SNR, these strong-amplitude probes will enable an accurate field estimation provided that they do not introduce significant non-linear terms into it.

\begin{figure}
   \resizebox{\hsize}{!}{\includegraphics{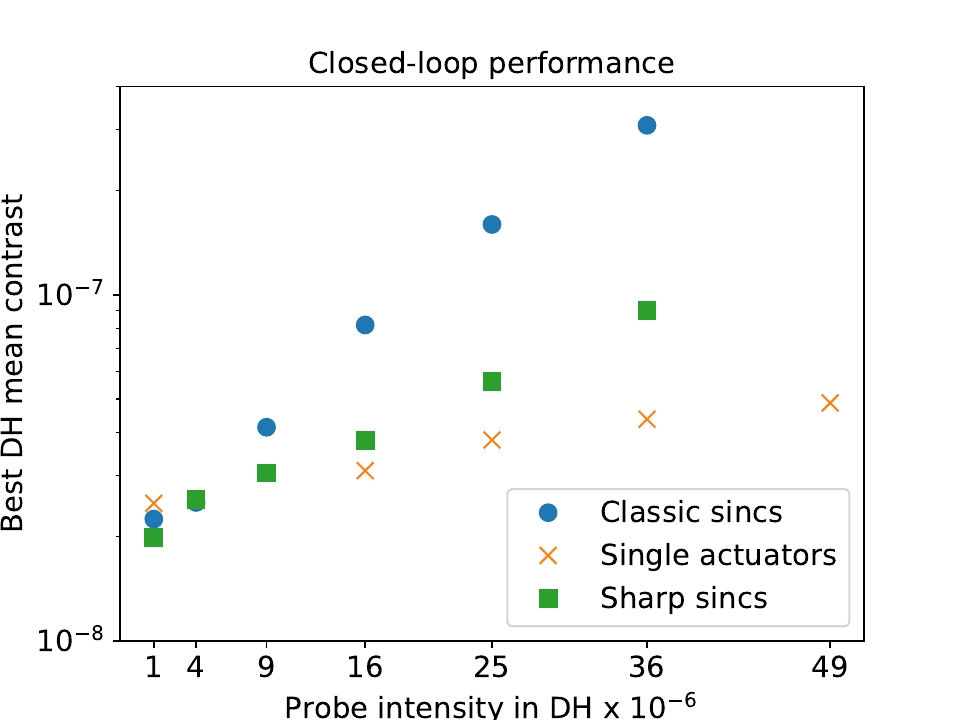}}
   \caption[Best performance per probe shape on hardware] 
   {\label{fig:best_performance} 
    Closed-loop performance comparison of different probe shapes, measured in experiments on THD2, as a function of probe intensity in the DH, scaled by $10^{-6}$. The contrast was measured between 3--9~$\lambda/D$. The plot shows the best achieved DH mean contrast for three shapes of probes: classic sinc-sinc-sine probes (blue circles), single-actuator probes (orange crosses), and sharp sinc probes (green squares). Single-actuator probes maintain consistent performance up to higher probe intensities with minimal contrast degradation, while classic sinc probes degrade significantly at higher intensities. Sharp sinc probes perform in between the other two, showing better resilience than classic sinc probes but not as robust as single-actuator probes. No DH loop has been recorded for single-actuator probes scaled up $\times$2 and $\times$3 in amplitude.
    The slightly unequal starting points at the initial amplitude come from the level of contrast uncertainty originating in drifts and environmental variations of the testbed, which is on the order of $10^{-8}$ for the mean contrast.}
   \end{figure}

Figure~\ref{fig:best_performance} shows the final mean DH contrast measured between 3--9~$\lambda/D$, achieved for a closed loop with each probe shape at each amplitude.
The results show that single-actuator probes maintain high contrast with minimal degradation even when their amplitude is scaled up by a factor of 6--7. Conversely, sinc-sinc-sine probes perform significantly worse at higher amplitudes. Sharp sinc probes fall in between the two, performing better than sinc-sinc-sine probes but not as robustly as single-actuator probes. No DH loop has been recorded for single-actuator probes scaled up $\times$2 and $\times$3 in amplitude. In the following section, we explore the reasons for this different behavior between the probe types as we focus on the most extreme solutions, inspecting a probe amplitude of 36$\times 10^{-6}$ in more detail.

\subsection{Non-linear terms in the pairwise observable $\Delta I_{\mathrm{im},\,j}(x,y)$}
\label{subsec:nonlinear_terms_PW_observable}

\cite{Groff2016MethodsLimitationsFocal} explain that probe amplitudes must fall within an acceptable range. If the amplitude is too low, the modulation signal gets buried in noise and the low SNR in the observed intensities leads to inaccuracies in the electric field estimation. If the probe amplitude is too high, non-linear terms in the Taylor expansion of the electric field become significant, causing the linear approximation of Eq.~\ref{eq:E_pup_expansion} to break down. We first expand the exponential of~Eq.\,\ref{eq:E_pup_expansion} to higher orders:
\begin{equation}
\begin{aligned}
  E_{\mathrm{pup},\,j} =& E_{\mathrm{pup}}\,\exp{\left(i\,\Psi_j\right)}\\ 
  \simeq &E_{\mathrm{pup}}\,\left(1+i\,\Psi_j-\frac{1}{2}\,\Psi_j^2-\frac{i}{6}\,\Psi_j^3+\frac{1}{24}\,\Psi_j^4\right).
\end{aligned}
\end{equation}
The image-plane electric field of Eq.~\ref{eq:E_im} then becomes
\begin{equation}
\begin{aligned}
         E_{\mathrm{im},\,j} \simeq\ & E_{\mathrm{im}} + i\,\mathcal{C}\{E_{\mathrm{pup}}\,\Psi_j\}
         -\frac{1}{2}\,C\{E_{\mathrm{pup}}\,\Psi_j^2\}\\
         &-\frac{i}{6}\,C\{E_{\mathrm{pup}}\,\Psi_j^3\} + \frac{1}{24}\,C\{E_{\mathrm{pup}}\,\Psi_j^4\}.
\end{aligned}
\end{equation}
As a consequence, Eq.\,\ref{eq:PW-delta_i_b}, which expresses the measured intensity difference for a probe pair~$\pm\Psi_j$, changes to:
\begin{equation}
\begin{aligned} 
\Delta I_{\mathrm{im},\,j} \simeq &4\ \Re{\left\{i\,E_{\mathrm{im}}\, \mathcal{C}\{E_{\mathrm{pup}}\,\Psi_j\}^*\right\}}\\
&-2\,\Re{\left\{i\,C\{E_{\mathrm{pup}}\,\Psi_j\}\,C\{E_{\mathrm{pup}}\,\Psi_j^2\}^*\right\}}\\
&-\frac{2}{3}\,\Re{\left\{i\,E_{\mathrm{im}}\, C\{E_{\mathrm{pup}}\,\Psi_j^3\}^*\right\}}\\
&+\frac{1}{3}\,\Re{\left\{i\, C\{E_{\mathrm{pup}}\,\Psi_j^2\} \,C\{E_{\mathrm{pup}}\,\Psi_j^3\}^* \right\}}\\
&+\frac{1}{6}\,\Re{\left\{i\,C\{E_{\mathrm{pup}}\,\Psi_j\}\,C\{E_{\mathrm{pup}}\,\Psi_j^4\}^*\right\}}.
\label{eq:PW-delta_i_high_orders}
\end{aligned}
\end{equation}
We label the five terms on the right side of this equation as Term~I, II, III, IV, and V. These terms correspond to the five terms of Eq.~(85) in \citet[same order]{Groff2016MethodsLimitationsFocal}. Term~I represents the linear approximation obtained in~Eq.\,\ref{eq:PW-delta_i_b}.  Terms II and III are third-order terms in~$\Psi_j$, and Terms IV and V are fifth-order terms.

For small amplitudes, the linear approximation captured in Term~I predominates and the other terms can be neglected, making PW probing a valid algorithm to estimate the linearized electric field. However, in the presence of high probe amplitudes, these higher-order cross terms between the electric field and the probe fields become significant. When sufficiently strong, they lead to errors in the linear electric field reconstruction since they behave as a bias in the inversion of the observation matrix in Eq.~\ref{eq:PW-x-hat-estimate_b}.

\begin{figure*}
   \includegraphics[scale=0.88]{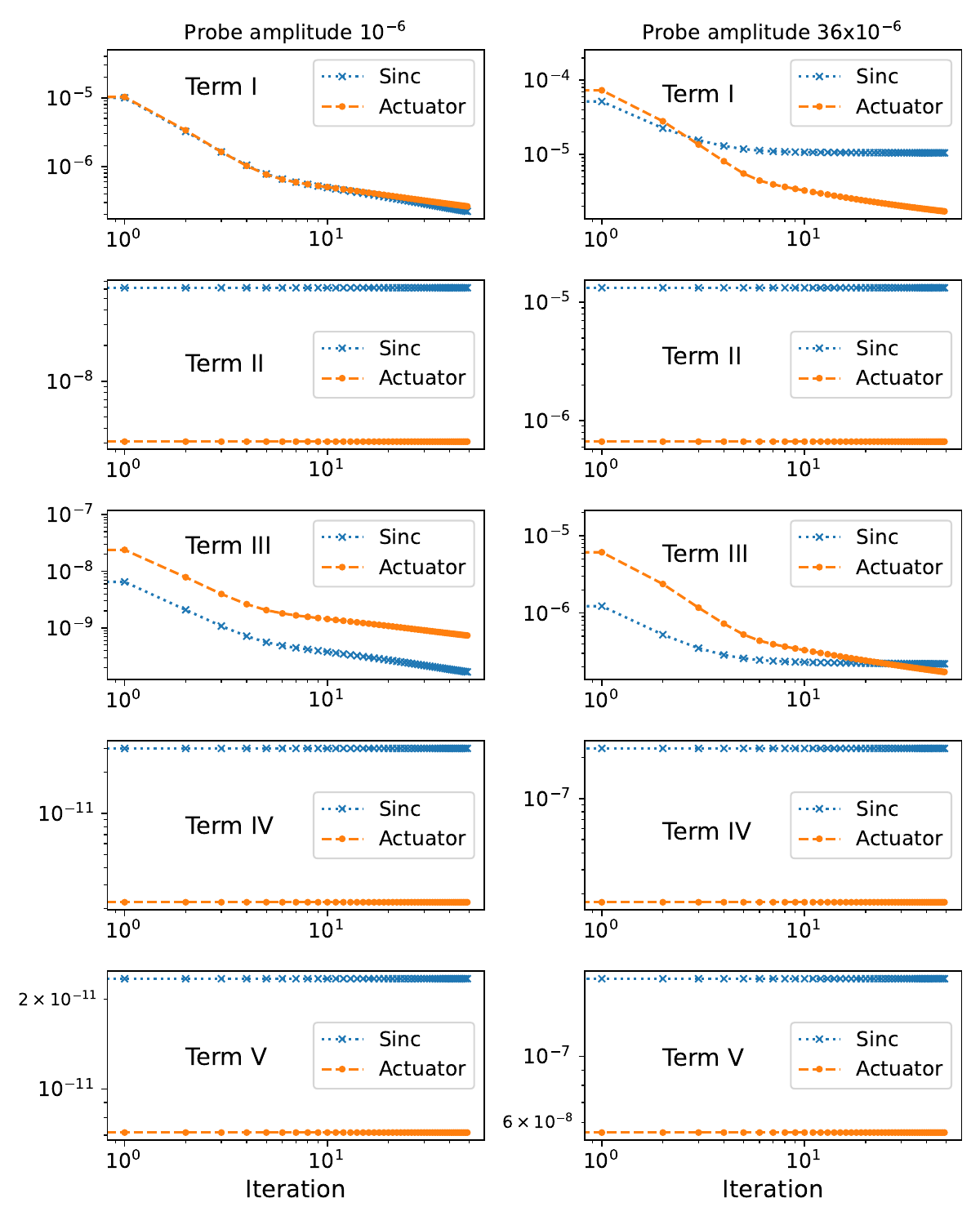}
   \caption[Individual terms per iteration] 
   {\label{fig:i_v_terms} 
    Evolution of the standard deviation for each term in Eq.~\ref{eq:PW-delta_i_high_orders} across the DH region during 50 iterations of a PW+EFC simulation. Results are shown for two shapes of probes: sinc-sinc-sine probes (blue crosses) and single-actuator probes (orange dashed lines), using probe amplitudes scaled to an added DH mean contrast of $10^{-6}$ (left column), and scaled up by a factor of 6, corresponding to an added DH mean contrast of $36\times10^{-6}$ (right column). Each plot corresponds to a different term: Term~I (linear approximation of $\Delta I_{\mathrm{im},\,j}$), Term~II (dominant non-linear term), Term~III, Term~IV, and Term~V. While Terms~II, IV, and V remain invariant across iterations, Term~I and Term~III depend on the aberrated electric field and decrease as the electric field improves. Notably, Term~II starts dominating with the classic sinc probes as probe amplitude increases, and Term~III decreases more rapidly for single-actuator probes compared to classic sinc probes.}
   \end{figure*}

Using numerical simulation, we calculated each of these five terms for classic sinc and single-actuator probes using our optical model. This was done throughout a PW + EFC run of 50 iterations, with all simulation parameters matching those in Sec.~\ref{subsec:pw_estimation_accuracy}.
To evaluate these term contributions to $\Delta I_{\mathrm{im},\,j}$, we computed the standard deviation of each term across the DH region.
Results for runs with two probe amplitudes are shown in Fig.~\ref{fig:i_v_terms}. The first case is a moderate probe amplitude adding $10^{-6}$ mean contrast in the DH, which we have confirmed to behave well in closed loop in Sec.~\ref{subsec:pw_estimation_accuracy}. The second case shows strong probes, scaled up by a factor of 6 from the moderate probes, leading to a relative DH mean contrast of $36\times10^{-6}$.

For the moderate probe amplitude in the left column of Fig.~\ref{fig:i_v_terms}, we observe how the linear Term~I is orders of magnitude stronger than all the calculated non-linear terms. As described in Eq.~\ref{eq:PW-delta_i_high_orders}, only Terms~I and III depend on the aberrated electric field, which makes them decrease over the course of the loop. Term~I stays dominant in all cases.

Between the constant Terms~II, IV and V, Term~II is the strongest one. In the case of the sinc-sinc-sine, it becomes larger than the linear Term~I in the strong probe case, shown in the right column of Fig.~\ref{fig:i_v_terms}. This has already been pointed out by \cite{Groff2016MethodsLimitationsFocal}. The key finding here is that in the regime of probes with strong amplitudes, Term~II is almost ten times higher for sinc-sinc-sine probes than for the single-actuators probes. Simulations (figures not shown in this paper) show that for the sharp sinc probes, the Terms~I and~II contributions lie between those of the classic sinc and single-actuator probes.
In future studies, we will explore other probes that obey Eq.~\ref{eq:necessary-condition_b} and minimize Term~II at the same time. Meanwhile, we conclude that the single-actuator probes perform better in terms of non-linearities than the sinc-sinc-sine probes.

\subsection{Fast iterations through blind estimation with strong probes at low flux}
\label{subsec:blind_correction}

Efficient and rapid iterations in HCI experiments are crucial for achieving deep contrasts while managing the constraints of instrument resources and time. There is technically no need to measure the unprobed images of a DH loop until a desired contrast level is reached. It is thus possible to run a loop ``blindly'', where contrast is estimated, but not directly measured. The probed images, however, must maintain sufficient SNR to enable accurate electric field estimation. This requirement becomes challenging under low-flux conditions, as the exposure times necessary for well-exposed images can be long when the applied probes are not very strong. By increasing the amplitude of probes, the required SNR can be reached or alternatively, exposure times can be shortened. However, probe amplitudes are inherently constrained by the linear regime of the PW probing algorithm. This section explores the application of strong probes in such ``blind'' loops under low-flux, demonstrating their potential to reduce exposure times and accelerate WFS\&C processes.

\begin{figure}
   \resizebox{\hsize}{!}{\includegraphics{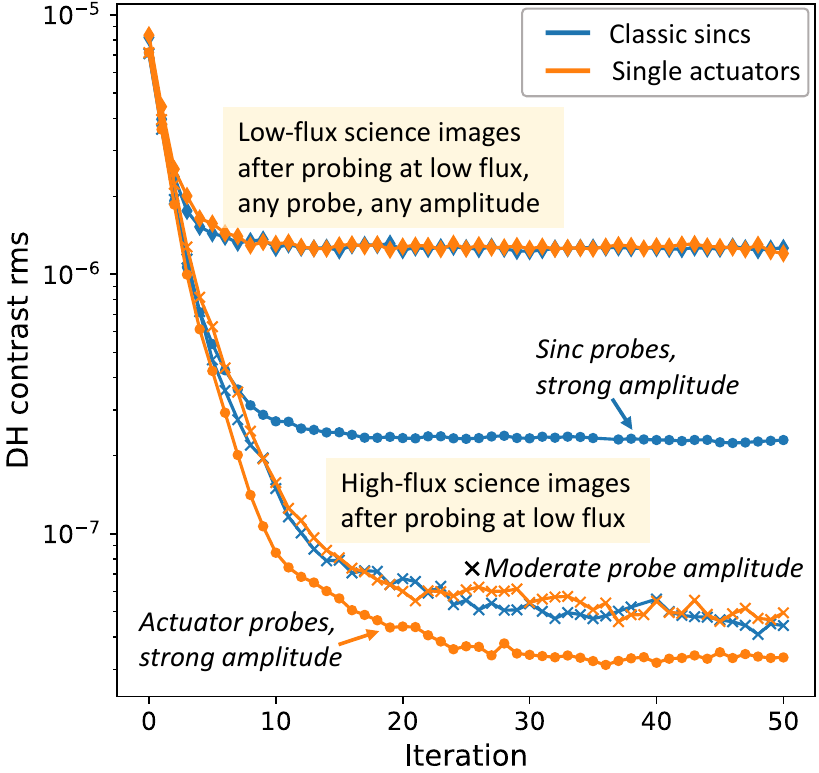}}
   \caption[Blind correction] 
   {\label{fig:blind_correction} 
    Evolution of the DH contrast rms as a function of iteration for different probe shapes and amplitudes estimated at low flux, and measured at high flux in experiments on THD2. The diamond markers indicate the noisy DH contrast measured during closed-loop control at low flux, where the contrast bottoms out at the detector's noise floor of $1.3\times10^{-6}$. Crosses and circles represent the effective DH contrast measured \textit{a posteriori} at high flux in open loop, corresponding to DM commands obtained from low-flux runs with moderate and strong probes, respectively. Classic sinc and single-actuator probes are compared, showing that strong-amplitude single-actuator probes achieve faster convergence to $5\times10^{-8}$ contrast compared to other configurations. The results demonstrate the advantage of strong probes in speeding up the correction process under low-flux, noisy conditions, particularly for fainter stars.}
   \end{figure}

\begin{figure}
   \resizebox{\hsize}{!}{\includegraphics{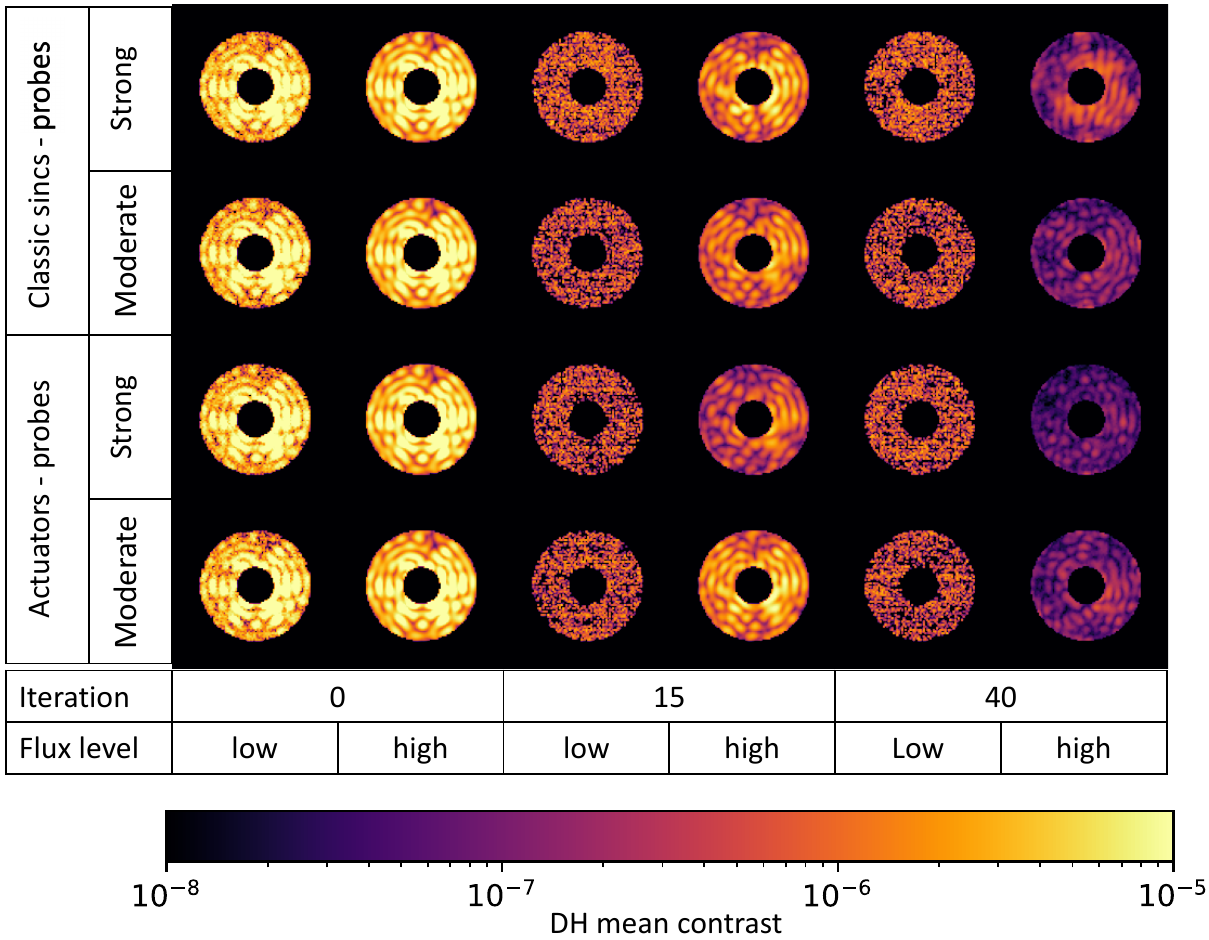}}
   \caption[Blind correction] 
   {\label{fig:blind_images} 
    Visual comparison of DH images obtained during the experiment for various probe shapes (classic sinc and single-actuator) and probe amplitudes (strong and moderate) under low- and high-flux conditions. Rows represent the probe configurations (classic sinc and single-actuator), with each configuration tested for strong and moderate probe amplitudes. Columns depict iterations at flux levels transitioning from low to high.}
   \end{figure}

Our findings in the preceding sections suggest that single-actuator probes allow for higher probe amplitudes compared to traditional sinc-sinc-sine probes. This permits the use of shorter exposure times or the use of fainter sources, effectively reducing the operational overhead of DH corrections. To validate this hypothesis, we conducted an experiment on the THD2 testbed, applying the sinc-sinc sine probes and single-actuator probes under low-flux conditions. The recorded correction commands were then applied in open loop at high flux to measure the effectively reached contrast. The results of this experiment are shown in Fig.~\ref{fig:blind_correction}, with corresponding DH images shown in Fig.~\ref{fig:blind_images}.

For all runs in this section, we operate in monochromatic light. EFC regularization is kept constant throughout, and the loop gain is set to 0.2. The probing experiments were preceded by an initial reference run of 200 iterations with moderate-amplitude single-actuator probes at high flux to establish a baseline for subsequent measurements. This is the high-flux, moderate-amplitude probing reference imitating a perfect baseline. We recorded its final coronagraphic image and associated DM commands for subsequent drift estimation during the following experiments. After each of them, this drift is measured below a $1.5\times10^{-8}$ contrast rms, which is below the experiment measurements hereafter.

From the same initial aberration level, we now close the loop at low flux, simulating a more realistic case of observation (see below) with limited SNR on the images. At this low flux level, the speckles became indistinguishable after a few iterations, necessitating strong probes to elevate the modulating signal above the read-out noise of the detector. As such, it is more representative to measure the standard deviation of contrast inside the DH region, as the mean would take positive or negative values not drawable in logarithmic scale. The diamond curves in the top of Fig.~\ref{fig:blind_images} show this standard deviation for the contrast in the DH for this moderate-amplitude, low flux case. Since this shows the unmodulated images, the data is the same in the strong-amplitude case at low flux, for any probe shape.

Under these low-flux conditions, we ran a closed loop both with moderate- ($10^{-6}$) and strong- ($36\times10^{-6}$) amplitude probes, with the single-actuator and the sinc-sinc-sine probes each, resulting in four loops. The strong-amplitude probes were scaled by a factor of six relative to the moderate-amplitude probes. Their measured DH mean contrast for a single-actuator and classic sinc probe are $4.0\times10^{-5}$ and $4.8\times10^{-5}$, respectively.
Due to the low flux, these loops do not directly measure the actual contrast but rely only on the estimations.
In each iteration of these four runs, we recorded the DM control commands and the coronagraphic images.

At the conclusion of all four runs, we increase the flux to record open-loop images using the DM commands recorded from each iteration in the previous low-flux experiments. This gives the images that would be recorded in each iteration during the blind correction (circles for strong probes and crosses for moderate probes in Fig.~\ref{fig:blind_images}).

The diamond curves show the noisy measured contrast during the closed-loop control at low flux. Since the SNR in these images is so low, this contrast quickly bottoms out at the noise level around $1.3\times10^{-6}$ rms, which corresponds to the readout noise per pixel of our detector. As expected, the contrast in the coronagraphic images is quickly dominated by this readout noise and cannot be used for planet detection.

To explain the curves recorded at high flux (circles and crosses), we first estimate the SNR of the measurement of the speckle intensity in the probe images. The associated noise term includes a read-out-noise per speckle, and photon noise contribution. With 4.63~pixels per $\lambda/D_{Lyot}$ in the images, we can estimate the read-out-noise per speckle to be about $3.2\times10^{-7}$. We estimated an SNR of 23 for the strong-amplitude probe images, and a SNR of 3.6, about 6 times lower, for the moderate-amplitude probe images. Therefore, both the moderate- and the strong-amplitude probes provide an SNR that is sufficient for an accurate estimation of the speckle intensity at low flux, and the subsequent EFC minimization. This is why the resulting coronagraphic contrast recorded at high flux using the ``blind estimation at low flux'' shows very good performance.

Like in Sec.~\ref{subsec:pw_estimation_accuracy}, we find that the moderate-amplitude probes perform equally well for single-actuator and classic sinc probes. We also find that the strong-amplitude sinc-sinc-sine probes have a limited performance, as explained in Sec.~\ref{subsec:nonlinear_terms_PW_observable}. Finally, the strong-amplitude single-actuator probes provide the best performance because the SNR of the PW probe images is significantly higher compared to moderate-amplitude probes (6 times higher), while not running into non-linear issues.

To compare these data with practical Roman operations, we estimate that the flux injected on the bench for each probed image corresponds to an .75~s (12~s) exposure time for a star of magnitude V=2 (V=5) in the band 1b of Roman ($\lambda_0$=575~nm, $\Delta\lambda$=18.98~nm, assuming an instrumental transmission of 47\%, \citep{Seo2018HybridLyotCoronagraph}).
Note that PW probing requires 6 images when using 3 probe pairs (here we assume that we do not record the contrast image since it is too noisy) for each wavelength, at three wavelengths for polychromatic corrections. Thus, each broadband iteration would last about 14~s for V=2, and 3.5~min for V=5. As shown in Fig. \ref{fig:blind_correction}, to reach a $5\times10^{-8}$ contrast, it takes half the time (17 iterations) with strong-amplitude actuator probes (orange circles) compared to moderate-amplitude single-actuator probes (37 iterations, orange crosses), while strong-amplitude classic sinc probes never reach this contrast (blue circles). With very high flux and the same EFC configuration and moderate probes, the number of iterations required to reach such level on the bench is 11. Thus, while the convergence speed in terms of iterations is not as good as in this perfect case, the strong probes really help speeding up the correction. The real advantage shows itself when comparing the time spent on the respective number of iterations to reach a $5\times10^{-8}$ contrast: For a magnitude 2 star, assuming the polychromatic correction is as efficient as the monochromatic one (remembering that the loop gain and control matrix regularization is far from optimal here), the strong-amplitude single-actuator probes (SNR of 23 for the PW estimation) reach this level after 17$\times14$~s (4 minutes) instead of $37\times 14$~s (8.6 minutes) in the moderate-amplitude probe case (SNR of 3.6 for the PW estimation).

\section{Discussion}
\label{sec:discussion}

The findings presented in this study demonstrate the potential advantages of single-actuator probes over traditional sinc-sinc-sine probes in PW probing for focal-plane wavefront sensing. Single-actuator probes exhibit reduced non-linearities at higher probe amplitudes, as described in Sec.~\ref{subsec:nonlinear_terms_PW_observable}, allowing for faster convergence at low flux while maintaining comparable performance in achieving contrast, which we demonstrated in laboratory experiments in Sec.~\ref{subsec:blind_correction}. There might potentially be differences at deeper contrasts than achieved here, as highlighted by \cite{Groff2016MethodsLimitationsFocal} in the context of maximum control step limits imposed by probe amplitude.

Single-actuator probes are conceptually closer to an ideal Dirac delta function, resulting in more uniform modulation of focal-plane pixels, even at higher orders. We suspect this property being advantageous for optimizing Term II in Eq.~\ref{eq:PW-delta_i_high_orders}, the dominant non-linear contribution in PW probing at high amplitudes, although more work is required to confirm this thesis. Our results show that Term II is significantly lower for single-actuator probes compared to sinc-sinc-sine probes. Term III, the next strongest non-linear term, decreases faster during iterations when using single-actuator probes.

One potential approach to mitigating the impact of non-linearities is to calculate Term II in simulations and subtract it from the $\Delta I_{\mathrm{im},\,j}(x,y)$ measurement prior to estimation. While this compensation strategy could improve estimation accuracy, optimizing Term II at its source remains desirable to further enhance overall performance. Term III, by contrast, is naturally eliminated when using the non-linear PW formalism as shown by \cite{Groff2016MethodsLimitationsFocal}, leaving higher-order terms negligible in practical scenarios.

Despite these advances, there may be yet better probe shapes to discover, particularly configurations where the focal-plane response of the square of a probe is orthogonal to its first-order response. Such designs would theoretically result in zero contribution to the inversion of the observation matrix, $H$, in Eq.~\ref{eq:PW-x-hat-estimate_b}.

Single-actuator probes are straightforward to implement. They require minimal calibration, involving only the selection of three actuators and a probe amplitude. However, effective implementation depends on understanding the DM influence function in the context of a specific instrument. For example, while our results suggest these findings should generalize across different DM configurations, including the one used in Roman CGI, further studies are needed to confirm this applicability.

Roman CGI employs DMs that differ from those on THD2 \citep{Krist2023EndToEndNumericalModeling}, raising the question of whether these differences could impact the conclusions of this paper. Our numerical simulations demonstrate that the linear regime extension of the probe amplitude is a consequence of fundamental mathematical and optical principles, suggesting these findings should hold true regardless of specific DM configurations. A direct comparison of the DM actuator influence functions suggests that the difference should be minor to none: The inter-actuator coupling (height of the influence function at the center of the first neighboring actuator) is $\sim$12\% for a Boston DM on THD2 \citep{Mazoyer2014DeformableMirrorInterferometric} and $\sim$17\% for a Xinetics DM on CGI \citep{Krist2023EndToEndNumericalModeling}. Scaled to the number of visible actuators in the pupil (28 for THD2 and 47 for CGI), the difference in ratio between influence function and pupil diameter is only $\sim$3\%. Follow-up investigations should focus on testing single-actuator probes with a full optical model of Roman CGI \citep{Krist2023EndToEndNumericalModeling,Zhou2023RomanCoronagraphHOWFSC}.

Working with strong-amplitude probes offers a clear benefit in terms of speed. It speeds up the DH creation by the same factor the flux is reduced compared to a well-exposed case, which reduces operational overhead, liberating time for science observations. A more accurate estimation of the effectively saved time needs inputs from official CGI models. Moreover, the CGI team has identified model mismatch as a primary factor contributing to slow convergence and the observed contrast floor. Simplifying the probe design, as demonstrated with single-actuator probes, could help mitigate such mismatches by reducing the complexity of the optical model.

Blind DH correction as presented and demonstrated in laboratory experiments on THD2 in Sec.~\ref{subsec:blind_correction} offers an additional advantage when working with faint sources. It could be applied directly on the science target, potentially eliminating the need for telescope slews. Even in scenarios where a separate target is required, the faster DH digging enabled by shorter exposure times would still yield significant efficiency gains. However, practical application of these methods on Roman specifically requires further investigation. The ground-in-the-loop time might dominate the WFS\&C process for CGI, particularly since the current plan might restrict operations to a single iteration per session. Nevertheless, deploying single-actuator probes on CGI -- or even better-performing probes -- would provide an opportunity to gain fundamental insights into PW probing.

On THD2, the closed-loop implementation was deliberately constrained to focus on investigating the behavior of PW probing. Specifically, we did not recalculate the control matrix with updated DM commands after each iteration, excluded high-order modes from the EFC regularization, and maintained a low loop gain throughout. These choices led to relatively slow convergence and a final DH contrast limited to $10^{-8}$ in standard deviation. While suboptimal for reaching deeper contrasts, these parameters provided a controlled environment to isolate and analyze the performance of different probing strategies, ensuring the robustness of our conclusions about single-actuator probes.

In future work, we aim to extend our findings to larger spectral bandwidths once the new broadband light source becomes available on THD2. An upcoming infrastructure and software upgrade \citep{Por2024catkit2} will further facilitate variable experimental setups, streamlining the investigation of advanced probing techniques. While the THD2 testbed operates in air and differs from CGI in its DM design, electronics, and environmental conditions, its similar contrast regime makes it a valuable proxy for understanding fundamental PW probing dynamics, particularly when complemented by numerical simulations.


\section{Conclusions}
\label{sec:conclusions}

This study demonstrates an optimization of pair-wise (PW) probing strategies for high-contrast imaging, specifically in the context of the Roman Coronagraphic Instrument (CGI). By exploring three distinct probe designs -- sinc-sinc-sine probes, single-actuator probes, and sharp sinc probes -- we examined their performance across critical parameters such as focal-plane modulation uniformity, non-linear effects at high amplitudes, and efficiency in achieving deep contrasts.

Our findings indicate that single-actuator probes offer a notable advantage over traditional sinc-sinc-sine probes. These probes exhibited reduced susceptibility to non-linear contributions in the observed intensity difference in PW probing, particularly at higher probe amplitudes. This characteristic enabled faster convergence to target contrasts without compromising performance. Sharp sinc probes showed intermediate performance, offering some benefits over sinc-sinc-sine probes but not achieving the robustness of single-actuator probes.

High-amplitude probing revealed important trade-offs. While increasing probe amplitude improved signal-to-noise ratios in low-flux scenarios and reduced exposure times, it introduced non-linearities that could degrade estimation accuracy. Single-actuator probes demonstrated superior resilience to these effects, maintaining effective performance even at amplitudes significantly higher than those suitable for sinc-sinc-sine probes. This makes them particularly advantageous for ``blind'' dark-hole digging in low-flux conditions where contrast is not directly measured, opening up a potentially feasible scenario for on-sky operations with fainter targets.

The integration of these findings into the broader context of Roman CGI operations is promising. Our results underscore the potential for single-actuator probes to enhance WFS\&C strategies, reducing operational overhead and improving efficiency. The scalability of these conclusions to the flight configuration of Roman CGI will need to be confirmed in further work with the mission-specific modeling tools and instrument parameters. The optical and mathematical root of our findings, though, highlight the relevance of this work for future missions, including the Habitable Worlds Observatory.

The THD2 testbed has proven to be a highly relevant platform for investigating these advanced WFS\&C techniques. Operating in a contrast regime comparable to that of Roman CGI, THD2 enables systematic exploration of probe designs, control strategies, and non-linear effects in a controlled laboratory environment. Its flexible configuration and capabilities to simulate space-like conditions make it an invaluable tool for bridging the gap between theoretical modeling and flight operations. This study exemplifies how insights gained on THD2 can directly inform and optimize the operational strategies for missions like Roman and beyond.


\begin{acknowledgements}

This research was developed in Python, an open source programming language. We report use the of the Numpy \citep{Oliphant2006numpy, vanderWalt2011NumPyArrayStructure}, Matplotlib \citep{Hunter2007matplotlib, matplotlib_v3.3.3}, and Astropy \citep{AstropyCollaboration2013, AstropyCollaboration2018, Astropy2018zenodo} packages. This research made use of HCIPy, an open-source object-oriented framework written in Python for performing end-to-end simulations of high-contrast imaging instruments \citep{Por2018HighContrastImaging}.

Some hardware components were calibrated by student B. Roman. I.L. is supported by the European Space Agency (ESA) under the tender number TDE-TEC-MMO AO/1-11613/23/NL/AR in the context of the ``SUPPPPRESS'' project. The development of the THD2 test bench was partly supported by Centre National d'\'Etudes Spatiales (CNES) through R\&D fundings R-S14/SU-002-068, R-S17/SU-0002-068, R-S19/SU-0002-105.
 
\end{acknowledgements}

\bibliographystyle{aa}
\bibliography{2024thd2roman}

\label{LastPage}
\end{document}